\documentclass[submission,copyright,creativecommons]{eptcs}
\providecommand{\event}{Linearity-TLLA 2020} 
\usepackage{breakurl}             
\usepackage{underscore}           
\usepackage{amsmath}
\usepackage{amssymb,graphicx}
\usepackage{proof}
\usepackage{tikz}
\usetikzlibrary{trees}
\newcommand{\rarrow}{\rotatebox[origin=c]{180}{\,$\multimap$\,}}
\newcommand{\id}{\mathit{id}}
\newcommand{\BB}{\mathbf{B}}
\newcommand{\II}{\mathbf{I}}
\newcommand{\bfC}{\mathbf{C}}
\newcommand{\CC}{\mathbf{C}}

\newcommand{\comment}[1]{}
\newcommand{\sem}[1]{[\![#1]\!]}
\newcommand{\ssigma}{{\boldsymbol\sigma}}
\newcommand{\idline}[3]{\put(#1,#2){\line(1,0){#3}}}
\newcounter{doubleV}
\newcounter{XplusH}
\newcounter{YplusV}
\newcounter{XplusHH}
\newcounter{YplusVV}

\newcount\mycount
\newcommand{\trace}[4]
    {\setcounter{doubleV}{#4}
     \addtocounter{doubleV}{#4}
     \setcounter{XplusH}{#1}
     \addtocounter{XplusH}{#3}
     \setcounter{YplusV}{#2}
     \addtocounter{YplusV}{#4}
     \put(#1,#2){\oval(\value{doubleV},\value{doubleV})[l]}
     \put(#1,\value{YplusV}){\line(1,0){#3}}
     \put(\value{XplusH},#2){\oval(\value{doubleV},\value{doubleV})[r]}
     }
\newcommand{\braid}[3]
     {\setcounter{XplusH}{#1}
      \addtocounter{XplusH}{#3}
     \setcounter{YplusV}{#2}
     \addtocounter{YplusV}{#3}
     \put(#1,#2){\line(1,1){#3}}
     \mycount=#3 \divide \mycount by 2
     \setcounter{XplusHH}{#1}
     \addtocounter{XplusHH}{\mycount}
     \addtocounter{XplusHH}{-2}
     \setcounter{YplusVV}{#2}
     \addtocounter{YplusVV}{\mycount}
     \addtocounter{YplusVV}{2}
     \qbezier(#1,\value{YplusV})(#1,\value{YplusV})(\value{XplusHH},\value{YplusVV})
     \addtocounter{XplusHH}{4}
     \addtocounter{YplusVV}{-4}
     \qbezier(\value{XplusH},#2)(\value{XplusH},#2)(\value{XplusHH},\value{YplusVV})
}

\newcommand{\braidInv}[3]
     {\setcounter{XplusH}{#1}
      \addtocounter{XplusH}{#3}
     \setcounter{YplusV}{#2}
     \addtocounter{YplusV}{#3}
     \put(#1,\value{YplusV}){\line(1,-1){#3}}
     \mycount=#3 \divide \mycount by 2
     \setcounter{XplusHH}{#1}
     \addtocounter{XplusHH}{\mycount}
     \addtocounter{XplusHH}{-2}
     \setcounter{YplusVV}{#2}
     \addtocounter{YplusVV}{\mycount}
     \addtocounter{YplusVV}{-2}
     \qbezier(#1,#2)(#1,#2)(\value{XplusHH},\value{YplusVV})
     \addtocounter{XplusHH}{4}
     \addtocounter{YplusVV}{4}
     \qbezier(\value{XplusH},\value{YplusV})(\value{XplusH},\value{YplusV})(\value{XplusHH},\value{YplusVV})
}
\newtheorem{theorem}{Theorem}

\newtheorem{lemma}{Lemma}
\newtheorem{proposition}{Proposition}

\newtheorem{remark}{Remark}
\newtheorem{example}{Example}

\newcommand{\celticbraidcolor}[6]{%
\put(#1,#2){%
\makebox(0,0)[bl]{
\begin{picture}(25,40)(0,0) \thicklines
#3
\qbezier(0,40)(5,40)(10,35)
\qbezier(10,35)(10,35)(23,22)
#6
\qbezier(0,0)(5,0)(10,5)
\qbezier(15,10)(15,10)(25,20)
#4
\qbezier(0,20)(0,20)(9,29)
\qbezier(15,35)(20,40)(25,40)
#5
\qbezier(3,17)(3,17)(15,5)
\qbezier(15,5)(20,0)(25,0)
\end{picture}}}}

\newcommand{\Cp}{\mathbf{C}^+}
\newcommand{\Cm}{\mathbf{C}^-}

\title{A Braided Lambda Calculus}
\author{Masahito Hasegawa
\institute{Research Institute for Mathematical Sciences\\Kyoto University\\Kyoto, Japan}
\email{hassei@kurims.kyoto-u.ac.jp}
}
\def\titlerunning{A Braided Lambda Calculus}
\def\authorrunning{M. Hasegawa}
\begin{document}
\maketitle
\begin{abstract}
We present an untyped linear lambda calculus with braids,
the corresponding combinatory logic, and the semantic models
given by crossed $G$-sets.
\end{abstract}
\section{Introduction}

\paragraph{Braids}

A {\em braid with $n$-strands} \cite{Art25,Art47,KT08} is $n$ copies of the interval $[0,1]$
smoothly embedded in the cube $[-\frac{1}{2},\frac{1}{2}]\times[0,1]\times[0,1]$ 
(Figure \ref{fig:braids}) such that
\begin{itemize}
\item each $t\in[0,1]$ is mapped to a point in the plane $\{(x,y,z)~|~z=t\}$
\item the end points $0\in[0,1]$ are sent to the $n$ points 
$\{(0,\frac{k}{n-1},0)~|~k=0,\dots,n-1\}$
\item the end points $1\in[0,1]$ are sent to the $n$ points 
$\{(0,\frac{k}{n-1},1)~|~k=0,\dots,n-1\}$
\end{itemize}
Two braids are identified if there is a continuous deformation between them 
preserving the boundaries (the ambient isotopy).
It is well-known that braids (modulo ambient isotopy)
can be identified with their projections to a
plane modulo Reidemeister moves, and also with the elements of 
the braid group: 
$$
\begin{array}{ll}
& \{\mbox{braids of $n$-strands}\}/\mbox{ambient isotopy}\\
\cong & \{\mbox{braid diagrams of $n$-strands}\}/\mbox{Reidemeister moves}\\
\cong & \mbox{Braid group $B_n$}
\end{array}
$$

\begin{figure}
\newsavebox{\braidP}
\savebox{\braidP}{%
\begin{picture}(20,20)(0,0)\thicklines
\qbezier(0,0)(0,7)(10,10)
\qbezier(20,20)(20,13)(10,10)
\qbezier(20,0)(20,7)(13,9)
\qbezier(0,20)(0,13)(7,11)
\end{picture}}

\newsavebox{\braidN}
\savebox{\braidN}{%
\begin{picture}(20,20)(0,0)\thicklines
\qbezier(20,0)(20,7)(10,10)
\qbezier(0,20)(0,13)(10,10)
\qbezier(20,0)(20,7)(13,9)
\qbezier(0,20)(0,13)(7,11)
\end{picture}}

\begin{center}
\begin{picture}(300,100)
{
\put(25,25){\vector(1,0){75}}
\put(25,25){\vector(0,1){75}}
\put(40,40){\vector(-1,-1){35}}
}
\put(15,75){\line(1,0){60}}
\put(15,15){\line(0,1){60}}
\put(15,15){\line(1,0){60}}
\put(75,15){\line(0,1){60}}
\put(35,95){\line(1,0){60}}
\put(35,35){\line(0,1){60}}
\put(35,35){\line(1,0){60}}
\put(95,35){\line(0,1){60}}
\put(15,75){\line(1,1){20}}
\put(75,15){\line(1,1){20}}
\put(75,75){\line(1,1){20}}
\put(25,85){\line(1,0){60}}
\put(85,25){\line(0,1){60}}
\put(25,25){\makebox(0,0){$\bullet$}}
\put(45,25){\makebox(0,0){$\bullet$}}
\put(65,25){\makebox(0,0){$\bullet$}}
\put(85,25){\makebox(0,0){$\bullet$}}
\put(25,85){\makebox(0,0){$\bullet$}}
\put(45,85){\makebox(0,0){$\bullet$}}
\put(65,85){\makebox(0,0){$\bullet$}}
\put(85,85){\makebox(0,0){$\bullet$}}
\put(2,12){\makebox(0,0){$x$}}
\put(102,30){\makebox(0,0){$y$}}
\put(20,100){\makebox(0,0){$z$}}
\put(90,22){\makebox(0,0){$1$}}
\put(20,90){\makebox(0,0){$1$}}
\put(67,18){\makebox(0,0){$\frac{2}{3}$}}
\put(47,18){\makebox(0,0){$\frac{1}{3}$}}
\put(20,27){\makebox(0,0){$0$}}
\put(10,22){\makebox(0,0){$\frac{1}{2}$}}
\put(28,40){\makebox(0,0){$-\frac{1}{2}$}}
\thicklines
\qbezier(25,25)(45,30)(45,50)
\qbezier(45,50)(45,60)(40,70)
\qbezier(37,73)(35,75)(25,85)
\qbezier(45,25)(40,30)(40,30)
\qbezier(38,33)(30,50)(30,60)
\qbezier(30,60)(30,73)(65,85)
\qbezier(65,25)(80,40)(80,50)
\qbezier(80,50)(80,55)(75,62)
\qbezier(71,65)(60,75)(52,78)
\qbezier(49,81)(45,85)(45,85)
\qbezier(85,25)(77,33)(77,33)
\qbezier(73,37)(65,60)(85,85)
\thinlines
\put(125,85){\line(1,0){60}}
\put(125,25){\line(1,0){60}}
\put(125,25){\makebox(0,0){$\bullet$}}
\put(145,25){\makebox(0,0){$\bullet$}}
\put(165,25){\makebox(0,0){$\bullet$}}
\put(185,25){\makebox(0,0){$\bullet$}}
\put(125,85){\makebox(0,0){$\bullet$}}
\put(145,85){\makebox(0,0){$\bullet$}}
\put(165,85){\makebox(0,0){$\bullet$}}
\put(185,85){\makebox(0,0){$\bullet$}}
\put(125,25){\usebox{\braidP}}
\put(125,45){\usebox{\braidP}}
\put(145,65){\usebox{\braidP}}
\put(165,25){\usebox{\braidP}}
\put(165,45){\usebox{\braidP}}
\thicklines
\qbezier(125,65)(125,65)(125,85)
\qbezier(185,65)(185,65)(185,85)
\put(285,50){\makebox(0,0){$\ssigma_1\ssigma_3^2\ssigma_1\ssigma_2$}}
\put(50,-5){\makebox(0,0){braid with 4 strands}}
\put(155,-5){\makebox(0,0){braid diagram}}
\put(285,-5){\makebox(0,0){element of the braid group $B_4$}}
\end{picture}
\end{center}
\caption{Braids}
\label{fig:braids}
\end{figure}

\paragraph{A braided lambda calculus}

In this paper, 
we introduce an {\em untyped linear lambda calculus with braids}, in which every 
permutation/exchange of variables is realized by a braid. 
Thus, for a term $M$ with $n$ (ordered) free variables and a braid $s$
with $n$ strands, we introduce a term $[s]M$ in which the 
free variables are permutated by $s$:
$$
\infer[\mathsf{braid}]
{x_{s(1)},x_{s(2)},\dots,x_{s(n)}\vdash {[s]}M}
{x_1,x_2,\dots,x_n\vdash M & {s:\mbox{braid with $n$ strands}}}
$$

For instance, we have two braided $\mathbf{C}$-combinators 
$$
\mathbf{C}^+\equiv\lambda fxy.
\left[
\begin{picture}(30,20)(-5,-2)
\thicklines
\qbezier(0,0)(5,0)(10,5)
\qbezier(10,5)(15,10)(20,10)
\qbezier(0,10)(5,10)(7,8)
\qbezier(13,2)(15,0)(20,0)
\qbezier(0,-10)(0,-10)(20,-10)
\put(-5,10){\makebox(0,0){\footnotesize$y$}}
\put(-5,0){\makebox(0,0){\footnotesize$x$}}
\put(-5,-10){\makebox(0,0){\footnotesize$f$}}
\put(25,10){\makebox(0,0){\footnotesize$x$}}
\put(25,0){\makebox(0,0){\footnotesize$y$}}
\put(25,-10){\makebox(0,0){\footnotesize$f$}}
\end{picture}
\right]
(f\,y\,x)
~~~~~
\mathbf{C}^-\equiv\lambda fxy.
\left[
\begin{picture}(30,20)(-5,-2)
\thicklines
\qbezier(0,10)(5,10)(10,5)
\qbezier(10,5)(15,0)(20,0)
\qbezier(0,0)(5,0)(7,2)
\qbezier(13,8)(15,10)(20,10)
\qbezier(0,-10)(0,-10)(20,-10)
\put(-5,10){\makebox(0,0){\footnotesize$y$}}
\put(-5,0){\makebox(0,0){\footnotesize$x$}}
\put(-5,-10){\makebox(0,0){\footnotesize$f$}}
\put(25,10){\makebox(0,0){\footnotesize$x$}}
\put(25,0){\makebox(0,0){\footnotesize$y$}}
\put(25,-10){\makebox(0,0){\footnotesize$f$}}
\end{picture}
\right]
(f\,y\,x)
$$
which are  ``implementations'' of the standard $\mathbf{C}$-combinator $\lambda fxy.f\,y\,x$ using braids. The idea of realizing a braided calculus as a planar calculus 
enriched with explicit braids is not new, see for instance \cite{Fle03}.


\paragraph{Why braids?}

Braids do not play a serious role in most of the conventional computational models,
and for the time being this work is largely a mathematical exercise with no immediate
 application.
Nevertheless,  let us say a little bit more on the motivation of this work and its 
potential applications. 

Extensionally, permutations (symmetry/exchange) are used for 
swapping two data.
On the other hand, braids provide non-extensional information on 
{how to implement permutations in three dimensions}. If braids have
some computational meaning, it should be something about 
{low-level (intermediate) codes to be compiled in some 3D computational architectures}.
One such computational model allowing ``braids for implementation''
reading is {\em Topological Quantum Computation} \cite{Kit03}, where the topological
information of anyons in 3D space-time does matter;
we hope that this work will find some usage in this context.
In a larger perspective, this work forms part of our research project on 
relating low-level codes and low-dimensional topology via categorical machineries.

From a more abstract point of view, our braided lambda calculus and the corresponding
braided combinatory algebras are algebraic structures which can be
described in terms of {\em PROBs} ({\em pro}ducts and {\em b}raids categories), the braided
version of PROPs (products and permutations categories). 
It seems that the theory of PROBs is a sort of folklore and 
there are very few published works on it (cf. \cite{nlabprop,Ver17}); we expect that the 
braided lambda calculus serves as a good test case of PROBs, e.g. the 
treatment of substitutions in braided algebras.

\paragraph{Why linear?}
Our calculus is linear, as there is no non-trivial  braid in a non-linear setting.
When the tensor product is  cartesian, any braid $\sigma_{A,B}:A\times B\rightarrow B\times A$ is equal to
the symmetry 
$\langle\pi'_{A,B},\pi_{A,B}\rangle
:A\times B\rightarrow B\times A$
because 
\newsavebox{\braidPP}
\savebox{\braidPP}{%
\begin{picture}(20,20)(0,0)
\thicklines
\qbezier(0,0)(5,0)(10,5)
\qbezier(10,5)(15,10)(20,10)
\qbezier(0,10)(5,10)(7,8)
\qbezier(13,2)(15,0)(20,0)
\end{picture}
}
\newsavebox{\braidNN}
\savebox{\braidNN}{%
\begin{picture}(20,20)(0,0)
\thicklines
\qbezier(0,10)(5,10)(10,5)
\qbezier(10,5)(15,0)(20,0)
\qbezier(0,0)(5,0)(7,2)
\qbezier(13,8)(15,10)(20,10)
\end{picture}
}
\newsavebox{\deltabox}
\savebox{\deltabox}{%
\begin{picture}(20,30)(0,0)
\thicklines
\put(0,20){\line(2,1){20}}
\put(0,10){\line(0,1){10}}
\put(0,10){\line(2,-1){20}}
\put(20,0){\line(0,1){30}}
\put(10,15){\makebox(0,0){$\Delta$}}
\end{picture}
}%
\comment{%
\begin{center}
\begin{picture}(20,50)
\put(0,20){\usebox{\braidPP}}
\end{picture}
\begin{picture}(50,50)
\put(25,35){\makebox(0,0){\scriptsize$\langle\pi,\pi'\rangle\!=\!\id$}}
\put(25,25){\makebox(0,0){$=$}}
\end{picture}
\begin{picture}(60,50)
\thicklines
\put(0,20){\usebox{\braidPP}}
\put(20,10){\usebox{\deltabox}}
\idline{40}{10}{20}
\idline{40}{20}{10} \put(50,20){\makebox(0,0){$\bullet$}}
\idline{40}{30}{10} \put(50,30){\makebox(0,0){$\bullet$}}
\idline{40}{40}{20}
\end{picture}
\begin{picture}(60,50)
\put(30,35){\makebox(0,0){\scriptsize naturality of $\Delta$}}
\put(30,25){\makebox(0,0){$=$}}
\end{picture}
\begin{picture}(60,50)
\thicklines
\idline{0}{20}{10}
\idline{0}{30}{10}
\put(10,10){\usebox{\deltabox}}
\put(30,10){\usebox{\braidPP}}
\put(30,30){\usebox{\braidPP}}
\idline{50}{10}{10}
\put(50,20){\makebox(0,0){$\bullet$}}
\put(50,30){\makebox(0,0){$\bullet$}}
\idline{50}{40}{10}
\end{picture}
\begin{picture}(60,50)
\put(30,42){\makebox(0,0){\scriptsize naturality of $\sigma$}}
\put(30,35){\makebox(0,0){\scriptsize $\sigma_{1,A}=\id_A=\sigma_{A,1}$}}
\put(30,25){\makebox(0,0){$=$}}
\end{picture}
\begin{picture}(60,50)
\thicklines
\idline{0}{20}{10}
\idline{0}{30}{10}
\put(10,10){\usebox{\deltabox}}
\idline{30}{10}{10} \put(40,10){\makebox(0,0){$\bullet$}}
\idline{30}{20}{20}
\idline{30}{30}{20}
\idline{30}{40}{10} \put(40,40){\makebox(0,0){$\bullet$}}
\end{picture}
\end{center}
}%
$$
\begin{array}{rcll}
\sigma_{A,B} 
&=&
\langle\pi_{B,A},\pi'_{B,A}\rangle\circ\sigma_{A,B} & 
\langle\pi,\pi'\rangle=\mathit{id}\\
&=&
\langle\pi_{B,A}\circ\sigma_{A,B},\pi'_{B,A}\circ\sigma_{A,B}\rangle\\
&=&
\langle\pi'_{A,B},\pi_{A,B}\rangle
\end{array}
$$
where $\pi_{B,A}\circ\sigma_{A,B}=\pi'_{A,B}$ because
$$
\begin{array}{rcll}
\pi_{B,A}\circ\sigma_{A,B}
&=&
(\mathit{id}_B\times !_A)\circ\sigma_{A,B}\\
&=&
\sigma_{1,B}\circ (!_A\times\mathit{id}_B) & \mbox{naturality of $\sigma$}\\
&=&
\mathit{id}_B\circ\pi'_{A,B} & \sigma_{1,B}=\mathit{id}_B\\
&=&
\pi'_{A,B}
\end{array}
$$
Similarly $\pi'_{B,A}\circ\sigma_{A,B}=\pi_{A,B}$ holds.
Hence linearity is essential for studying a braided calculus in a
meaningful way. 

\paragraph{Why untyped?}
Our calculus is untyped. Compared to the typed case (including braided MLL \cite{Fle03}
and tensorial logic \cite{Mel18}), we have a simpler syntax 
and subtler, more challenging semantics - while the simply typed braided lambda calculus can be
modelled by any braided monoidal closed category, the untyped calculus requires
a reflexive object, which is hard to find in the well-known braided categories 
in TQFT \cite{Tur94}. We overcome this difficulty by using a braided relational
model constructed in our previous work \cite{Has12}. As far as we know, this is the
first non-trivial example of a reflexive object in a non-symmetric ribbon category.

\paragraph{Contributions}
Our contributions are summarized as follows.
\begin{itemize}
\item We formulate a braided lambda calculus whose syntax is a mild modification of the untyped linear lambda calculus with explicit braids (Section 2).
\item We introduce the corresponding combinatory logic and show 
the combinatory completeness which ensures that our combinatory logic is as 
expressive as the braided lambda calculus (Section 3).
\item We give categorical semantics given by reflexive objects in braided 
monoidal closed categories, and present some concrete models using crossed $G$-sets
(Section 4). 
\end{itemize}

\section{A Braided Lambda Calculus} 

\subsection{Syntax of the Calculus}

The untyped braided lambda calculus
is an extension of the planar lambda calculus (the linear lambda calculus with no
exchange)\footnote{%
In the literature, there are (at least) two different notions of ``planar lambda terms''.
Some authors employ the ``left'' abstraction rule (e.g. \cite{ZG15})
~~~\raise-2ex\hbox{\infer{\Gamma\vdash \lambda x.M}{x,\Gamma\vdash M}}~~~
whereas others (e.g. \cite{Abr07,Tom21}) use the ``right'' abstraction rule as 
we do in the present paper;
see \cite{ZG15} for some comparison.
Our choice has the advantage of preservation of planarity
under the $\beta\eta$-conversions, and allows simpler semantics
by reflexive objects in monoidal (right) closed categories.
}
with a rule for introducing braided terms.

$$
\infer[\mathrm{variable}]{x\vdash x}{}
~~~~~~
\infer[\mathrm{abstraction}]{\Gamma\vdash \lambda x.M}{\Gamma,x\vdash M}
~~~~~~
\infer[\mathrm{application}]{\Gamma,\Gamma'\vdash M\,N}{\Gamma\vdash M & \Gamma'\vdash N}
$$


$$
\infer[\mathrm{braid}]
{x_{s(1)},x_{s(2)},\dots,x_{s(n)}\vdash [s]M}
{x_1,x_2,\dots,x_n\vdash M & s:\mbox{braid with $n$ strands}}
$$
where $s(i)$ denotes the outcome of applying the permutation
on $\{1,2,\dots,n\}$ induced by $s$ to $i$.
Formally, a braid with $n$ strands will be an element of the braid group $B_n$,
and the braided term $[s]M$ is the result of the group action of $B_n$ on terms with
$n$ free variables. 
However, for readability, we might present braids graphically,
often with labels indicating the correspondence to variables.

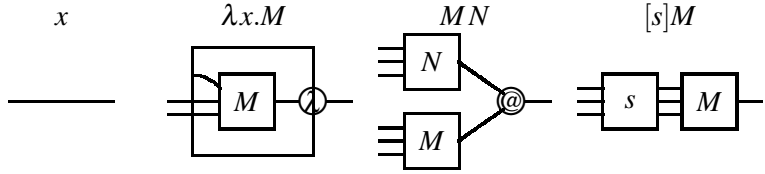
\begin{figure}[t]
$$
\begin{array}{cccc}
x & 
\lambda x.M & 
M\,N &
[s]M
\\
\begin{picture}(50,50)\thicklines
\idline{5}{25}{40}
\end{picture}
&
\begin{picture}(75,50)(-10,0)\thicklines
\put(5,5){\framebox(45,40){}}
\put(15,15){\framebox(20,20){$M$}}
\idline{-5}{20}{20}
\idline{-5}{25}{20}
\qbezier(5,35)(10,35)(15,30)
\put(50,25){\circle{10}}
\put(50,25){\makebox(0,0){$\lambda$}}
\idline{35}{25}{10}
\idline{55}{25}{10}
\end{picture}
&
\begin{picture}(65,50)\thicklines
\put(10,30){\framebox(20,20){$N$}}
\put(10,0){\framebox(20,20){$M$}}
\idline{0}{35}{10}
\idline{0}{40}{10}
\idline{0}{45}{10}
\idline{0}{5}{10}
\idline{0}{10}{10}
\idline{0}{15}{10}
\put(50,25){\circle{10}}
\put(50,25){\makebox(0,0){$@$}}
\idline{55}{25}{10}
\qbezier(30,40)(30,40)(47,28)
\qbezier(30,10)(30,10)(47,22)
\end{picture}
&
\begin{picture}(70,50)(5,0)\thicklines
\put(15,15){\framebox(20,20){$s$}}
\put(45,15){\framebox(20,20){$M$}}
\idline{5}{20}{10}
\idline{5}{25}{10}
\idline{5}{30}{10}
\idline{35}{20}{10}
\idline{35}{25}{10}
\idline{35}{30}{10}
\idline{65}{25}{10}
\end{picture}
\end{array}
$$
\caption{Term graphs}
\label{fig:termgraphs}
\end{figure}

It will be helpful to look at {\em term graphs} corresponding to 
 terms (Figure \ref{fig:termgraphs}), especially when discussing
the equational theory of the braided lambda calculus.

\begin{example}[braided C-combinator]
\comment{
There are infinite variations of the $\bfC$ combinator
in the braided lambda calculus. The most basic ones are:
$$
\mathbf{C}^+=\lambda fxy.
\left[
\begin{picture}(30,20)(-5,-2)
\thicklines
\qbezier(0,0)(5,0)(10,5)
\qbezier(10,5)(15,10)(20,10)
\qbezier(0,10)(5,10)(7,8)
\qbezier(13,2)(15,0)(20,0)
\qbezier(0,-10)(0,-10)(20,-10)
\put(-5,10){\makebox(0,0){\footnotesize$y$}}
\put(-5,0){\makebox(0,0){\footnotesize$x$}}
\put(-5,-10){\makebox(0,0){\footnotesize$f$}}
\put(25,10){\makebox(0,0){\footnotesize$x$}}
\put(25,0){\makebox(0,0){\footnotesize$y$}}
\put(25,-10){\makebox(0,0){\footnotesize$f$}}
\end{picture}
\right]
(f\,y\,x)
~~~~~
\mathbf{C}^-=\lambda fxy.
\left[
\begin{picture}(30,20)(-5,-2)
\thicklines
\qbezier(0,10)(5,10)(10,5)
\qbezier(10,5)(15,0)(20,0)
\qbezier(0,0)(5,0)(7,2)
\qbezier(13,8)(15,10)(20,10)
\qbezier(0,-10)(0,-10)(20,-10)
\put(-5,10){\makebox(0,0){\footnotesize$y$}}
\put(-5,0){\makebox(0,0){\footnotesize$x$}}
\put(-5,-10){\makebox(0,0){\footnotesize$f$}}
\put(25,10){\makebox(0,0){\footnotesize$x$}}
\put(25,0){\makebox(0,0){\footnotesize$y$}}
\put(25,-10){\makebox(0,0){\footnotesize$f$}}
\end{picture}
\right]
(f\,y\,x)
$$
}
The derivation of the combinator $\mathbf{C}^+$ in the introduction is
$$
\infer
{\vdash \lambda fxy.
\left[
s
\right]
(f\,y\,x)}
{\infer
{f\vdash \lambda xy.[s](f\,y\,x)}
{\infer
{f,x\vdash \lambda y.[s](f\,y\,x)}
{
\infer{
f,x,y\vdash
\left[
s
\right]
(f\,y\,x)}
{\infer{
 f,y,x\vdash f\,y\,x}
{\infer{f,y\vdash f\,y}{\infer{f\vdash f}{} & \infer{y\vdash y}{}} & \infer{x\vdash x}{}}&
s=~
\unitlength=.25mm
\begin{picture}(30,20)(-5,-10)
\thicklines
\qbezier(0,0)(5,0)(10,5)
\qbezier(10,5)(15,10)(20,10)
\qbezier(0,10)(5,10)(7,8)
\qbezier(13,2)(15,0)(20,0)
\qbezier(0,-10)(0,-10)(20,-10)
\put(-5,10){\makebox(0,0){\footnotesize$y$}}
\put(-5,0){\makebox(0,0){\footnotesize$x$}}
\put(-5,-10){\makebox(0,0){\footnotesize$f$}}
\put(25,10){\makebox(0,0){\footnotesize$x$}}
\put(25,0){\makebox(0,0){\footnotesize$y$}}
\put(25,-10){\makebox(0,0){\footnotesize$f$}}
\end{picture}
}}
}}
$$
and the term graph corresponding to $\CC^+$ is
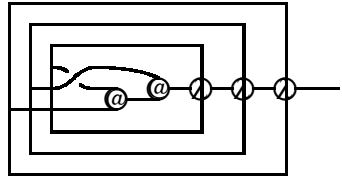
\begin{center}
\unitlength=.8pt
\begin{picture}(170,100)
\thicklines
\put(10,10){\framebox(130,80){}}
\put(140,50){\circle{10}}
\put(140,50){\makebox(0,0){$\lambda$}}
\put(20,20){\framebox(100,60){}}
\put(120,50){\circle{10}}
\put(120,50){\makebox(0,0){$\lambda$}}
\put(30,30){\framebox(70,40){}}
\put(100,50){\circle{10}}
\put(100,50){\makebox(0,0){$\lambda$}}
\qbezier(30,50)(35,50)(40,55)
\qbezier(40,55)(45,60)(50,60) \qbezier(50,60)(70,60)(80,55)
\qbezier(30,60)(35,60)(37,58)
\qbezier(43,52)(45,50)(50,50)
\qbezier(50,50)(50,50)(60,50)
\qbezier(20,50)(30,50)(30,50)
\qbezier(10,40)(30,40)(60,40)
\put(60,45){\circle{10}}
\put(60,45){\makebox(0,0){$@$}} 
\qbezier(65,45)(65,45)(80,45)
\put(80,50){\circle{10}}
\put(80,50){\makebox(0,0){$@$}}
\qbezier(85,50)(85,50)(95,50)
\qbezier(105,50)(105,50)(115,50)
\qbezier(125,50)(125,50)(135,50)
\qbezier(145,50)(145,50)(165,50)
\end{picture}
\end{center}
\end{example}

\begin{remark}[Contexts are redundant]
In the braided lambda calculus, the context 
is always uniquely determined by the term, thus redundant.
Given a braided lambda term $M$, we define the {\em list} $\mathrm{cxt}(M)$
of free variables in $M$ as follows:
$\mathrm{cxt}(x)=x$,~
$\mathrm{cxt}(M\,N)=\mathrm{cxt}(M),\mathrm{cxt}(N)$,~
$\mathrm{cxt}(\lambda x.M)=\Gamma$ where $\mathrm{cxt}(M)=\Gamma,x$,
and 
$\mathrm{cxt}([s]M)=s(\mathrm{cxt}(M))$
where $s(x_1,\dots,x_n)=x_{s(1)},\dots,x_{s(n)}$.
\comment{
\begin{itemize}
\item $\mathrm{cxt}(x)=x$.
\item $\mathrm{cxt}(M\,N)=\mathrm{cxt}(M),\mathrm{cxt}(N)$.
\item $\mathrm{cxt}(\lambda x.M)=\Gamma$ where $\mathrm{cxt}(M)=\Gamma,x$.
\item $\mathrm{cxt}([s]M)=s(\mathrm{cxt}(M))$, 
where $s(x_1,\dots,x_n)=x_{s(1)},\dots,x_{s(n)}$.
\end{itemize}
}
It follows that $\Gamma\vdash M$ iff $\mathrm{cxt}(M)=\Gamma$. 
Hence the context of a braided term is unique: 
if both $\Gamma\vdash M$ and $\Gamma'\vdash M$ are derivable,
then $\Gamma$ is identical to  $\Gamma'$.
\end{remark}
 
\subsection{Equational Theory} \label{subsec:eqational-theory}

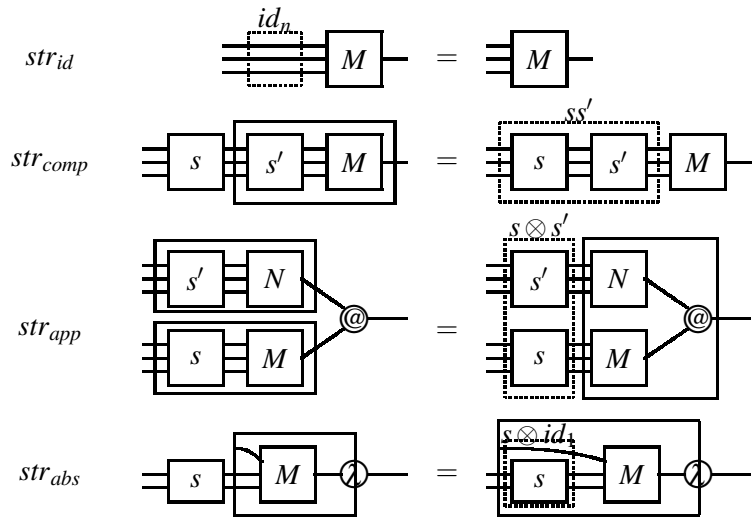
\begin{figure}
$$
\begin{array}{lrcl}
\begin{picture}(30,30)
\put(5,15){\makebox(0,0){$\mathit{str}_{\id}$}}
\thicklines
\end{picture}
&
\begin{picture}(70,30)
\thicklines
\put(10,5){\dashbox(20,20){}}
\put(20,30){\makebox(0,0){$\id_n$}}
\put(40,5){\framebox(20,20){$M$}}
\idline{0}{10}{40}
\idline{0}{15}{40}
\idline{0}{20}{40}
\idline{60}{15}{10}
\end{picture}
&
\begin{picture}(10,30)
\put(5,15){\makebox(0,0){$=$}}
\end{picture}
&
\begin{picture}(40,30)
\thicklines
\put(10,5){\framebox(20,20){$M$}}
\idline{0}{10}{10}
\idline{0}{15}{10}
\idline{0}{20}{10}
\idline{30}{15}{10}
\end{picture}
\\
\begin{picture}(10,30)
\put(5,15){\makebox(0,0){$\mathit{str}_{\mathit{comp}}$}}
\thicklines
\end{picture}
&
\begin{picture}(100,35)
\thicklines
\put(10,5){\framebox(20,20){$s$}}
\put(40,5){\framebox(20,20){$s'$}}
\put(70,5){\framebox(20,20){$M$}}
\idline{0}{10}{10}
\idline{0}{15}{10}
\idline{0}{20}{10}
\idline{30}{10}{10}
\idline{30}{15}{10}
\idline{30}{20}{10}
\idline{60}{10}{10}
\idline{60}{15}{10}
\idline{60}{20}{10}
\idline{90}{15}{10}
\put(35,0){\framebox(60,30){}}
\end{picture}
&
\begin{picture}(10,30)
\put(5,15){\makebox(0,0){$=$}}
\end{picture}
&
\begin{picture}(100,35)
\thicklines
\put(10,5){\framebox(20,20){$s$}}
\put(40,5){\framebox(20,20){$s'$}}
\put(70,5){\framebox(20,20){$M$}}
\idline{0}{10}{10}
\idline{0}{15}{10}
\idline{0}{20}{10}
\idline{30}{10}{10}
\idline{30}{15}{10}
\idline{30}{20}{10}
\idline{60}{10}{10}
\idline{60}{15}{10}
\idline{60}{20}{10}
\idline{90}{15}{10}
\put(5,0){\dashbox(60,30){}}
\put(35,35){\makebox(0,0){$ss'$}}
\end{picture}
\\
\begin{picture}(10,60)
\put(5,25){\makebox(0,0){$\mathit{str}_{\mathit{app}}$}}
\thicklines
\end{picture}
&
\begin{picture}(100,70)
\thicklines
\put(10,5){\framebox(20,20){$s$}}
\put(10,35){\framebox(20,20){$s'$}}
\put(40,5){\framebox(20,20){$M$}}
\put(40,35){\framebox(20,20){$N$}}
\put(80,30){\circle{10}}
\put(80,30){\makebox(0,0){$@$}}
\idline{0}{10}{10}
\idline{0}{15}{10}
\idline{0}{20}{10}
\idline{0}{40}{10}
\idline{0}{45}{10}
\idline{0}{50}{10}
\idline{30}{10}{10}
\idline{30}{15}{10}
\idline{30}{20}{10}
\idline{30}{40}{10}
\idline{30}{45}{10}
\idline{30}{50}{10}
\qbezier(60,15)(60,15)(76,26)
\qbezier(60,45)(60,45)(76,34)
\idline{85}{30}{15}
\put(5,2){\framebox(60,26){}}
\put(5,33){\framebox(60,26){}}
\end{picture}
&
\begin{picture}(10,60)
\put(5,25){\makebox(0,0){$=$}}
\thicklines
\end{picture}
&
\begin{picture}(100,70)
\thicklines
\put(10,5){\framebox(20,20){$s$}}
\put(10,35){\framebox(20,20){$s'$}}
\put(40,5){\framebox(20,20){$M$}}
\put(40,35){\framebox(20,20){$N$}}
\put(80,30){\circle{10}}
\put(80,30){\makebox(0,0){$@$}}
\idline{0}{10}{10}
\idline{0}{15}{10}
\idline{0}{20}{10}
\idline{0}{40}{10}
\idline{0}{45}{10}
\idline{0}{50}{10}
\idline{30}{10}{10}
\idline{30}{15}{10}
\idline{30}{20}{10}
\idline{30}{40}{10}
\idline{30}{45}{10}
\idline{30}{50}{10}
\qbezier(60,15)(60,15)(76,26)
\qbezier(60,45)(60,45)(76,34)
\idline{85}{30}{15}
\put(7,0){\dashbox(26,60){}}
\put(37,0){\framebox(50,60){}}
\put(20,65){\makebox(0,0){$s\otimes s'$}}
\end{picture}
\\
\begin{picture}(10,35)
\put(5,15){\makebox(0,0){$\mathit{str}_{\mathit{abs}}$}}
\thicklines
\end{picture}
&
\begin{picture}(100,40)
\thicklines
\put(45,5){\framebox(20,20){$M$}}
\put(10,5){\framebox(20,15){$s$}}
\put(80,15){\circle{10}}
\put(80,15){\makebox(0,0){$\lambda$}}
\idline{0}{10}{10}
\idline{0}{15}{10}
\idline{30}{10}{15}
\idline{30}{15}{15}
\qbezier(65,15)(65,15)(75,15)
\idline{85}{15}{15}
\put(35,0){\framebox(45,30){}}
\qbezier(35,25)(40,25)(45,20)
\end{picture}
&
\begin{picture}(10,35)
\put(5,15){\makebox(0,0){$=$}}
\thicklines
\end{picture}
&
\begin{picture}(100,40)
\thicklines
\put(45,5){\framebox(20,20){$M$}}
\put(10,5){\framebox(20,15){$s$}}
\put(80,15){\circle{10}}
\put(80,15){\makebox(0,0){$\lambda$}}
\idline{0}{10}{10}
\idline{0}{15}{10}
\idline{30}{10}{15}
\idline{30}{15}{15}
\qbezier(65,15)(65,15)(75,15)
\idline{85}{15}{15}
\put(5,0){\framebox(75,35){}}
\qbezier(5,25)(30,25)(45,20)
\put(7,3){\dashbox(26,25){}}
\put(20,30){\makebox(0,0){$s\otimes\id_1$}}
\end{picture}
\end{array}
$$
\caption{Structural axioms}
\label{fig:structuralaxioms}
\end{figure}

The $\beta\eta$-theory of the braided lambda calculus
has the usual $\beta\eta$ axioms plus 
structural axioms for braids:
$$
\begin{array}{lrcl}
\beta & (\lambda x.M)\,N &=& M[x:=N]\\
\eta & \lambda x.M\,x &=& M\\
\mathit{str}_{\id}& [\id_n]M &=& M ~~~~~~(M~\mathrm{has}~n~\mathrm{free~variables})\\
\mathit{str}_{\mathit{comp}} & [s]([s']M) &=& [ss']M\\
\mathit{str}_{\mathit{app}}& ([s]M)\,([s']N) &=& [s\otimes s'](M\,N)\\
\mathit{str}_{\mathit{abs}}& [s](\lambda x.M) &=& \lambda x.[s\otimes \id_1]M
\end{array}
$$
where $\id_n$ stands for the trivial braid with $n$ strands (the unit element $e$ of the braid group $B_n$), $ss'$ is the composition of
$s$ and $s'$ while $s\otimes s'$ the parallel composition
(Figure \ref{fig:operations-braids}).
The structural axioms identify two terms when 
they have the same underlying term graph
(Figure \ref{fig:structuralaxioms}).  
It might be worth pointing out that our calculus has some resemblance to
the calculi with explicit substitutions \cite{ACCL91}: 
braids can be thought as special substitutions (enriched with some extra information).

In the $\beta$ rule, the substitution $M[x:=N]$ means 
replacing the (unique) free variable $x$ in $M$ by $N$ and also $x$-labelled
strings occuring in
braids in $M$ by $\Gamma$-strings where $\Gamma\vdash N$. (When $N$ contains
no free variable, all $x$-strings are removed.)
This informal definition can be justified if we look at the corresponding term 
graphs: intuitively,
$$
\left(
\left[
\begin{picture}(32,12)(-6,2)
\thicklines
{
\qbezier(0,0)(5,0)(10,5)
\qbezier(10,5)(15,10)(20,10)}%
\qbezier(0,10)(5,10)(7,8)
\qbezier(13,2)(15,0)(20,0)
\put(-5,10){\makebox(0,0){\footnotesize$y$}}
\put(-5,0){\makebox(0,0){\footnotesize$x$}}
\put(25,10){\makebox(0,0){\footnotesize$x$}}
\put(25,0){\makebox(0,0){\footnotesize$y$}}
\end{picture}
\right]
(y\,{ x})\right)[{ x}:={ (x_1\,x_2)}]
~~~\mbox{should be}~~~ 
\left[
\begin{picture}(32,20)(-6,-2)
\thicklines
{
\qbezier(0,0)(5,0)(10,5)
\qbezier(10,5)(15,10)(20,10)
\qbezier(0,-10)(5,-10)(10,-5)
\qbezier(10,-5)(15,0)(20,0)}%
\qbezier(0,10)(4,10)(6,7)
\qbezier(9,1.5)(10,0)(11,-1.5)
\qbezier(14,-7)(16,-10)(20,-10)
\put(-5,10){\makebox(0,0){\footnotesize$y$}}
\put(-5,0){\makebox(0,0){\footnotesize$x_2$}}
\put(-5,-10){\makebox(0,0){\footnotesize$x_1$}}
\put(25,10){\makebox(0,0){\footnotesize$x_2$}}
\put(25,0){\makebox(0,0){\footnotesize$x_1$}}
\put(25,-10){\makebox(0,0){\footnotesize$y$}}
\end{picture}
\right]
(y\,{(x_1\,x_2)})
$$%
because they express the same term graph
(modulo continuous deformation):
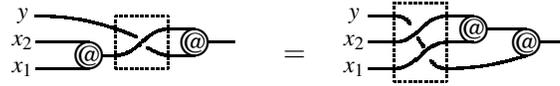
\begin{center}
\begin{picture}(75,30)
\thicklines
\put(30,5){\dashbox(20,20){}}
{
\idline{0}{15}{20}
\idline{0}{5}{20}}%
\put(20,10){\circle{10}}
\put(20,10){\makebox(0,0){@}}
\qbezier(0,25)(20,25)(38,17)
{
\idline{25}{10}{5}
\qbezier(30,10)(35,10)(40,15)
\qbezier(40,15)(45,20)(50,20)
\idline{50}{20}{10}}%
\qbezier(42,13)(47,10)(50,10)
\idline{50}{10}{10}
\put(60,15){\circle{10}}
\put(60,15){\makebox(0,0){@}}
\idline{65}{15}{10}
\put(-5,25){\makebox(0,0){\footnotesize$y$}}
\put(-5,15){\makebox(0,0){\footnotesize$x_2$}}
\put(-5,5){\makebox(0,0){\footnotesize$x_1$}}
\end{picture}
\begin{picture}(40,30)
\put(20,10){\makebox(0,0){$=$}}
\end{picture}
\begin{picture}(80,30)(-15,-15)
\thicklines
\put(0,-15){\dashbox(20,30){}}
\idline{-10}{10}{10}
{
\idline{-10}{0}{10}
\idline{-10}{-10}{10}
\qbezier(0,0)(5,0)(10,5)
\qbezier(10,5)(15,10)(20,10)
\qbezier(0,-10)(5,-10)(10,-5)
\qbezier(10,-5)(15,0)(20,0)
\idline{20}{10}{10}
\idline{20}{0}{10}}%
\qbezier(0,10)(4,10)(6,7)
\qbezier(9,1.5)(10,0)(11,-1.5)
\qbezier(14,-7)(16,-10)(20,-10)
\qbezier(20,-10)(35,-10)(50,-5)
\put(30,5){\circle{10}}
\put(30,5){\makebox(0,0){@}}
{\idline{35}{5}{15}}%
\put(50,0){\circle{10}}
\put(50,0){\makebox(0,0){@}}
\idline{55}{0}{10}
\put(-15,10){\makebox(0,0){\footnotesize$y$}}
\put(-15,0){\makebox(0,0){\footnotesize$x_2$}}
\put(-15,-10){\makebox(0,0){\footnotesize$x_1$}}
\end{picture}
\end{center}
Similarly, 
$$
\left(
\left[
\begin{picture}(32,12)(-6,2)
\thicklines
{
\qbezier(0,0)(5,0)(10,5)
\qbezier(10,5)(15,10)(20,10)}%
\qbezier(0,10)(5,10)(7,8)
\qbezier(13,2)(15,0)(20,0)
\put(-5,10){\makebox(0,0){\footnotesize$y$}}
\put(-5,0){\makebox(0,0){\footnotesize$x$}}
\put(25,10){\makebox(0,0){\footnotesize$x$}}
\put(25,0){\makebox(0,0){\footnotesize$y$}}
\end{picture}
\right]
(y\,x)\right)[x:=\lambda z.z]
~~~~\mbox{should be}~~~ ~
\left[
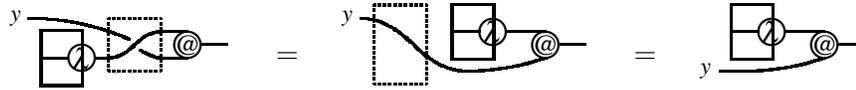
\begin{picture}(33,5)(-6,-2)
\thicklines
\qbezier(0,0)(0,0)(20,0)
\put(-5,0){\makebox(0,0){\footnotesize$y$}}
\put(25,0){\makebox(0,0){\footnotesize$y$}}
\end{picture}
\right]
(y\,(\lambda z.z))
~=_{\mathit{str}_{\id}}~
y\,(\lambda z.z).
$$
\begin{center}
\begin{picture}(75,30)
\thicklines
\put(30,5){\dashbox(20,20){}}
\put(20,10){\circle{10}}
\put(20,10){\makebox(0,0){$\lambda$}}
\idline{5}{10}{10}
\put(5,0){\framebox(15,20){}}
\qbezier(0,25)(20,25)(38,17)
{
\idline{25}{10}{5}
\qbezier(30,10)(35,10)(40,15)
\qbezier(40,15)(45,20)(50,20)
\idline{50}{20}{10}}%
\qbezier(42,13)(47,10)(50,10)
\idline{50}{10}{10}
\put(60,15){\circle{10}}
\put(60,15){\makebox(0,0){@}}
\idline{65}{15}{10}
\put(-5,25){\makebox(0,0){\footnotesize$y$}}
\end{picture}
\begin{picture}(40,30)
\put(20,10){\makebox(0,0){$=$}}
\end{picture}
\begin{picture}(90,30)(-25,-15)
\thicklines
\put(-15,-15){\dashbox(20,30){}}
\qbezier(-20,10)(-10,10)(0,0) \qbezier(0,0)(10,-10)(20,-10)
\qbezier(20,-10)(35,-10)(50,-5)
\put(30,5){\circle{10}}
\put(30,5){\makebox(0,0){$\lambda$}}
\idline{15}{5}{10}
\put(15,-5){\framebox(15,20){}}
{\idline{35}{5}{15}}%
\put(50,0){\circle{10}}
\put(50,0){\makebox(0,0){@}}
\idline{55}{0}{10}
\put(-25,10){\makebox(0,0){\footnotesize$y$}}
\end{picture}
\begin{picture}(40,30)
\put(20,10){\makebox(0,0){$=$}}
\end{picture}
\begin{picture}(80,30)(5,-15)
\thicklines
\idline{10}{-10}{10}
\qbezier(20,-10)(35,-10)(50,-5)
\put(30,5){\circle{10}}
\put(30,5){\makebox(0,0){$\lambda$}}
\idline{15}{5}{10}
\put(15,-5){\framebox(15,20){}}
{\idline{35}{5}{15}}%
\put(50,0){\circle{10}}
\put(50,0){\makebox(0,0){@}}
\idline{55}{0}{10}
\put(5,-10){\makebox(0,0){\footnotesize$y$}}
\end{picture}
\end{center}
Thus substitution is much subtler than one might first guess.
Below we discuss the formal definition of substitution, in which
braids are algebraically handled as elements of the braid group.

\begin{figure}[b]
\begin{center}
\begin{picture}(40,45)
\thicklines
\put(10,5){\dashbox(20,20){}}
\put(20,30){\makebox(0,0){$\id_3$}}
\idline{0}{10}{40}
\idline{0}{15}{40}
\idline{0}{20}{40}
\end{picture}
~~~~~~
\begin{picture}(70,45)
\thicklines
\put(10,5){\framebox(20,20){$s$}}
\put(40,5){\framebox(20,20){$s'$}}
\idline{0}{10}{10}
\idline{0}{15}{10}
\idline{0}{20}{10}
\idline{30}{10}{10}
\idline{30}{15}{10}
\idline{30}{20}{10}
\idline{60}{10}{10}
\idline{60}{15}{10}
\idline{60}{20}{10}
\put(5,0){\dashbox(60,30){}}
\put(35,35){\makebox(0,0){$ss'$}}
\end{picture}
~~~~~~
\begin{picture}(40,70)
\thicklines
\put(10,5){\framebox(20,20){$s$}}
\put(10,30){\framebox(20,20){$s'$}}
\idline{0}{10}{10}
\idline{0}{15}{10}
\idline{0}{20}{10}
\idline{0}{40}{10}
\idline{0}{45}{10}
\idline{0}{35}{10}
\idline{30}{10}{10}
\idline{30}{15}{10}
\idline{30}{20}{10}
\idline{30}{40}{10}
\idline{30}{45}{10}
\idline{30}{35}{10}
\put(7,0){\dashbox(26,55){}}
\put(20,60){\makebox(0,0){$s\otimes s'$}}
\end{picture}
\end{center}
\caption{Operations on braids}
\label{fig:operations-braids}
\end{figure}
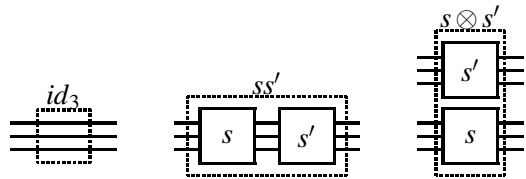

\begin{figure}
\begin{center}
\begin{tabular}{ccc}
\begin{picture}(110,80)(-10,0)\thicklines
\put(5,5){\framebox(45,40){}}
\put(15,15){\framebox(20,20){$M$}}
\idline{-5}{20}{20}
\idline{-5}{25}{20}
\qbezier(5,35)(10,35)(15,30)
\put(50,25){\circle{10}}
\put(50,25){\makebox(0,0){$\lambda$}}
\idline{35}{25}{10}
\qbezier(55,25)(55,25)(72,42)
\put(75,45){\circle{10}}
\put(75,45){\makebox(0,0){$@$}}
\put(15,60){\framebox(20,20){$N$}}
\idline{-5}{75}{20}
\idline{-5}{70}{20}
\idline{-5}{65}{20}
\qbezier(35,70)(35,70)(72,47)
\idline{80}{45}{20}
\end{picture}
&
\begin{picture}(20,80)(0,0)\thicklines
\put(10,35){\makebox(0,0){$=_\beta$}}
\end{picture}
&
\begin{picture}(90,80)(-10,0)\thicklines
\put(45,15){\framebox(20,20){$M$}}
\idline{-5}{20}{50}
\idline{-5}{25}{50}
\qbezier(25,40)(40,35)(45,30)
\put(5,30){\framebox(20,20){$N$}}
\idline{-5}{45}{10}
\idline{-5}{40}{10}
\idline{-5}{35}{10}
\idline{65}{25}{20}
\end{picture}
\\
\begin{picture}(110,80)(-10,0)\thicklines
\put(5,5){\framebox(65,50){}}
\put(15,15){\framebox(20,20){$M$}}
\idline{-5}{20}{20}
\idline{-5}{30}{20}
\put(50,30){\circle{10}}
\put(50,30){\makebox(0,0){$@$}}
\put(70,30){\circle{10}}
\put(70,30){\makebox(0,0){$\lambda$}}
\qbezier(5,45)(35,45)(47,32)
\qbezier(35,25)(35,25)(47,28)
\idline{55}{30}{10}
\idline{75}{30}{20}
\end{picture}
&
\begin{picture}(20,80)(0,0)\thicklines
\put(10,30){\makebox(0,0){$=_\eta$}}
\end{picture}
&
\begin{picture}(90,80)(-10,0)\thicklines
\put(15,15){\framebox(20,20){$M$}}
\idline{-5}{20}{20}
\idline{-5}{30}{20}
\idline{35}{25}{20}
\end{picture}
\end{tabular}
\end{center}
\caption{$\beta\eta$ axioms}
\label{fig:betaetaaxioms}
\end{figure}
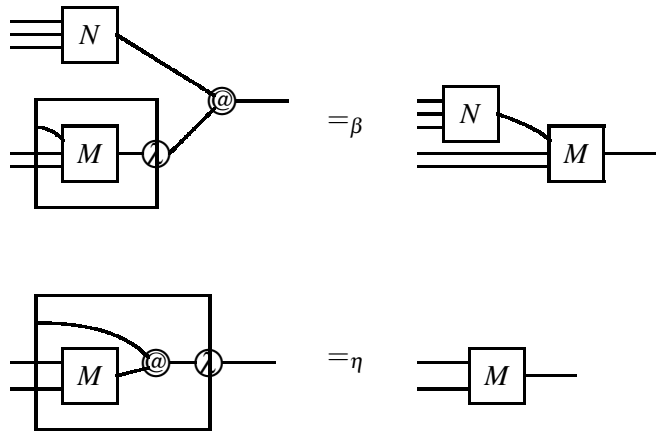

\subsection{Formal Treatment of Braids and Substitution}

\paragraph{The braid group}
Let $B_n$ be the Artin braid group \cite{Art25,Art47,KT08}
generated by $n-1$ generators
$\ssigma_1,\ssigma_2,\dots,\ssigma_{n-1}$ with relations
\begin{itemize}
\item $\ssigma_i\ssigma_j=\ssigma_j\ssigma_i$ for $1\leq i,j\leq n-1$ with $|i-j|\geq 2$, and
\item $\ssigma_i\ssigma_{i+1}\ssigma_i=\ssigma_{i+1}\ssigma_i\ssigma_{i+1}$ for $1\leq i\leq n-1$.
\end{itemize}
The following geometric reading in terms of braid diagrams may be
useful for understanding the behaviour of the generators $\ssigma_i$
and $\ssigma_i^{-1}$:
$$
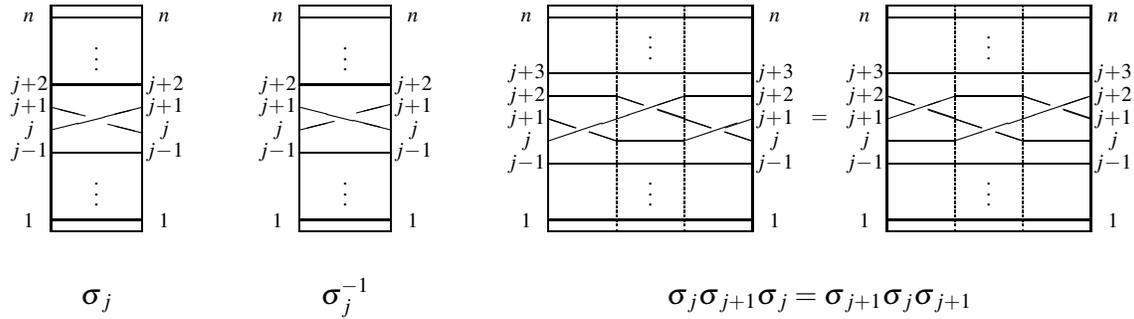
\unitlength=.3mm
\begin{array}{ccc}
~~~~~~
\begin{picture}(60,100)
\scriptsize
\put(10,0){\framebox(40,100){}}
\put(0,5){\makebox(0,0){$1$}}
\put(0,35){\makebox(0,0){$j\!-\!1$}}
\put(0,45){\makebox(0,0){$j$}}
\put(0,55){\makebox(0,0){$j\!+\!1$}}
\put(0,65){\makebox(0,0){$j\!+\!2$}}
\put(0,95){\makebox(0,0){$n$}}

\put(10,0){\framebox(40,100){}}
\put(60,5){\makebox(0,0){$1$}}
\put(60,35){\makebox(0,0){$j\!-\!1$}}
\put(60,45){\makebox(0,0){$j$}}
\put(60,55){\makebox(0,0){$j\!+\!1$}}
\put(60,65){\makebox(0,0){$j\!+\!2$}}
\put(60,95){\makebox(0,0){$n$}}

\put(10,5){\line(1,0){40}}
\put(10,35){\line(1,0){40}}

\put(10,45){\line(4,1){40}}

\put(10,55){\line(4,-1){15}}
\put(35,47.5){\line(4,-1){15}}
\put(10,65){\line(1,0){40}}
\put(10,95){\line(1,0){40}}

\put(30,20){\makebox(0,0){$\vdots$}}
\put(30,80){\makebox(0,0){$\vdots$}}
\end{picture}
~~~~~~
&
~~~~~~
\begin{picture}(60,100)
\scriptsize
\put(10,0){\framebox(40,100){}}
\put(0,5){\makebox(0,0){$1$}}
\put(0,35){\makebox(0,0){$j\!-\!1$}}
\put(0,45){\makebox(0,0){$j$}}
\put(0,55){\makebox(0,0){$j\!+\!1$}}
\put(0,65){\makebox(0,0){$j\!+\!2$}}
\put(0,95){\makebox(0,0){$n$}}

\put(10,0){\framebox(40,100){}}
\put(60,5){\makebox(0,0){$1$}}
\put(60,35){\makebox(0,0){$j\!-\!1$}}
\put(60,45){\makebox(0,0){$j$}}
\put(60,55){\makebox(0,0){$j\!+\!1$}}
\put(60,65){\makebox(0,0){$j\!+\!2$}}
\put(60,95){\makebox(0,0){$n$}}

\put(10,5){\line(1,0){40}}
\put(10,35){\line(1,0){40}}

\put(10,55){\line(4,-1){40}}

\put(10,45){\line(4,1){15}}
\put(35,52.5){\line(4,1){15}}
\put(10,65){\line(1,0){40}}
\put(10,95){\line(1,0){40}}

\put(30,20){\makebox(0,0){$\vdots$}}
\put(30,80){\makebox(0,0){$\vdots$}}
\end{picture}
~~~~~~
&
~~~~~~
\begin{picture}(260,100)
\scriptsize
\put(10,0){\framebox(90,100){}}
\put(40,0){\dashbox(0,100){}}
\put(70,0){\dashbox(0,100){}}
\put(0,5){\makebox(0,0){$1$}}
\put(0,30){\makebox(0,0){$j\!-\!1$}}
\put(0,40){\makebox(0,0){$j$}}
\put(0,50){\makebox(0,0){$j\!+\!1$}}
\put(0,60){\makebox(0,0){$j\!+\!2$}}
\put(0,70){\makebox(0,0){$j\!+\!3$}}
\put(0,95){\makebox(0,0){$n$}}
\put(110,5){\makebox(0,0){$1$}}
\put(110,30){\makebox(0,0){$j\!-\!1$}}
\put(110,40){\makebox(0,0){$j$}}
\put(110,50){\makebox(0,0){$j\!+\!1$}}
\put(110,60){\makebox(0,0){$j\!+\!2$}}
\put(110,70){\makebox(0,0){$j\!+\!3$}}
\put(110,95){\makebox(0,0){$n$}}
\put(10,5){\line(1,0){90}}
\put(10,30){\line(1,0){90}}
\put(10,70){\line(1,0){90}}
\put(10,95){\line(1,0){90}}
\put(55,20){\makebox(0,0){$\vdots$}}
\put(55,85){\makebox(0,0){$\vdots$}}
\put(10,40){\line(3,1){60}}
\put(70,60){\line(1,0){30}}
\put(10,50){\line(3,-1){12}}
\put(40,40){\line(-3,1){12}}
\put(40,40){\line(1,0){30}}
\put(70,40){\line(3,1){30}}
\put(10,60){\line(1,0){30}}
\put(40,60){\line(3,-1){12}}
\put(70,50){\line(-3,1){12}}
\put(70,50){\line(3,-1){12}}
\put(100,40){\line(-3,1){12}}
\put(130,50){\makebox(0,0){$=$}}
\put(160,0){\framebox(90,100){}}
\put(190,0){\dashbox(0,100){}}
\put(220,0){\dashbox(0,100){}}
\put(150,5){\makebox(0,0){$1$}}
\put(150,30){\makebox(0,0){$j\!-\!1$}}
\put(150,40){\makebox(0,0){$j$}}
\put(150,50){\makebox(0,0){$j\!+\!1$}}
\put(150,60){\makebox(0,0){$j\!+\!2$}}
\put(150,70){\makebox(0,0){$j\!+\!3$}}
\put(150,95){\makebox(0,0){$n$}}
\put(260,5){\makebox(0,0){$1$}}
\put(260,30){\makebox(0,0){$j\!-\!1$}}
\put(260,40){\makebox(0,0){$j$}}
\put(260,50){\makebox(0,0){$j\!+\!1$}}
\put(260,60){\makebox(0,0){$j\!+\!2$}}
\put(260,70){\makebox(0,0){$j\!+\!3$}}
\put(260,95){\makebox(0,0){$n$}}
\put(160,5){\line(1,0){90}}
\put(160,30){\line(1,0){90}}
\put(160,70){\line(1,0){90}}
\put(160,95){\line(1,0){90}}
\put(205,20){\makebox(0,0){$\vdots$}}
\put(205,85){\makebox(0,0){$\vdots$}}
\put(160,40){\line(1,0){30}}
\put(190,40){\line(3,1){60}}
\put(160,50){\line(3,1){30}}
\put(190,60){\line(1,0){30}}
\put(220,60){\line(3,-1){12}}
\put(250,50){\line(-3,1){12}}
\put(160,60){\line(3,-1){12}}
\put(190,50){\line(-3,1){12}}
\put(190,50){\line(3,-1){12}}
\put(220,40){\line(-3,1){12}}
\put(220,40){\line(1,0){30}}
\end{picture}
~~~~~~
\\
\\
\ssigma_j
&
\ssigma_j^{-1}
&
\ssigma_j\ssigma_{j+1}\ssigma_j=\ssigma_{j+1}\ssigma_j\ssigma_{j+1}
\end{array}
$$
In the sequel we will denote the unit element ($\id_n$) of the braid group by $e$.

\begin{figure}
$$
\begin{array}{ccc}
~~
\begin{picture}(60,100)
\scriptsize
\put(10,0){\framebox(40,100){}}
\put(0,5){\makebox(0,0){$1$}}
\put(0,35){\makebox(0,0){$j\!-\!1$}}
\put(0,45){\makebox(0,0){$j$}}
\put(0,55){\makebox(0,0){$j\!+\!1$}}
\put(0,65){\makebox(0,0){$j\!+\!2$}}
\put(0,95){\makebox(0,0){$n$}}

\put(10,0){\framebox(40,100){}}
\put(60,5){\makebox(0,0){$1$}}
\put(60,35){\makebox(0,0){$j\!-\!1$}}
\put(60,45){\makebox(0,0){$j$}}
\put(60,55){\makebox(0,0){$j\!+\!1$}}
\put(60,65){\makebox(0,0){$j\!+\!2$}}
\put(60,95){\makebox(0,0){$n$}}

\put(10,5){\line(1,0){40}}
\put(10,35){\line(1,0){40}}
{\thicklines
\put(10,45){\line(4,1){40}}
}
\put(10,55){\line(4,-1){15}}
\put(35,47.5){\line(4,-1){15}}
\put(10,65){\line(1,0){40}}
\put(10,95){\line(1,0){40}}

\put(30,20){\makebox(0,0){$\vdots$}}
\put(30,80){\makebox(0,0){$\vdots$}}
\end{picture}
~~
&
~~
\begin{picture}(60,100)
\scriptsize
\put(10,0){\framebox(40,100){}}
\put(0,5){\makebox(0,0){$1$}}
\put(0,25){\makebox(0,0){$j\!-\!1$}}
\put(0,35){\makebox(0,0){$j$}}
\put(0,45){\makebox(0,0){$j\!+\!1$}}
\put(0,55){\makebox(0,0){$j\!+\!2$}}
\put(0,65){\makebox(0,0){$j\!+\!3$}}
\put(0,75){\makebox(0,0){$j\!+\!4$}}
\put(0,95){\makebox(0,0){$n\!+\!2$}}

\put(10,0){\framebox(40,100){}}
\put(60,5){\makebox(0,0){$1$}}
\put(60,25){\makebox(0,0){$j\!-\!1$}}
\put(60,35){\makebox(0,0){$j$}}
\put(60,45){\makebox(0,0){$j\!+\!1$}}
\put(60,55){\makebox(0,0){$j\!+\!2$}}
\put(60,65){\makebox(0,0){$j\!+\!3$}}
\put(60,75){\makebox(0,0){$j\!+\!4$}}
\put(60,95){\makebox(0,0){$n\!+\!2$}}

\put(10,5){\line(1,0){40}}
\put(10,25){\line(1,0){40}}
{\thicklines
\put(10,35){\line(4,1){40}}
\put(10,45){\line(4,1){40}}
\put(10,55){\line(4,1){40}}
}
\qbezier(10,65)(10,65)(18,59)
\qbezier(22,56)(22,56)(28,51.5)
\qbezier(32,48.5)(32,48.5)(38,44)
\qbezier(42,41)(42,41)(50,35)

\put(10,75){\line(1,0){40}}
\put(10,95){\line(1,0){40}}

\put(30,15){\makebox(0,0){$\vdots$}}
\put(30,85){\makebox(0,0){$\vdots$}}
\end{picture}
~~
&
~~
\begin{picture}(60,100)
\scriptsize
\put(10,0){\framebox(40,100){}}
\put(0,5){\makebox(0,0){$1$}}
\put(0,35){\makebox(0,0){$j\!-\!1$}}
\put(0,50){\makebox(0,0){$j$}}
\put(0,65){\makebox(0,0){$j\!+\!1$}}
\put(0,95){\makebox(0,0){$n\!-\!1$}}

\put(10,0){\framebox(40,100){}}
\put(60,5){\makebox(0,0){$1$}}
\put(60,35){\makebox(0,0){$j\!-\!1$}}
\put(60,50){\makebox(0,0){$j$}}
\put(60,65){\makebox(0,0){$j\!+\!1$}}
\put(60,95){\makebox(0,0){$n\!-\!1$}}

\put(10,5){\line(1,0){40}}
\put(10,35){\line(1,0){40}}
\put(10,50){\line(1,0){40}}
\put(10,65){\line(1,0){40}}
\put(10,95){\line(1,0){40}}

\put(30,20){\makebox(0,0){$\vdots$}}
\put(30,80){\makebox(0,0){$\vdots$}}
\end{picture}
~~
\\
\\
\ssigma_j
&
~~~~~~~~~\ssigma_j[j:=3]=\ssigma_{j+2}\ssigma_{j+1}\ssigma_j~~~~~~~~~
&
\ssigma_j[j:=0]=e(=\mathit{id}_{n-1})
\end{array}
$$
\caption{Substitution map}
\label{fig:substitutionmap}
\end{figure}
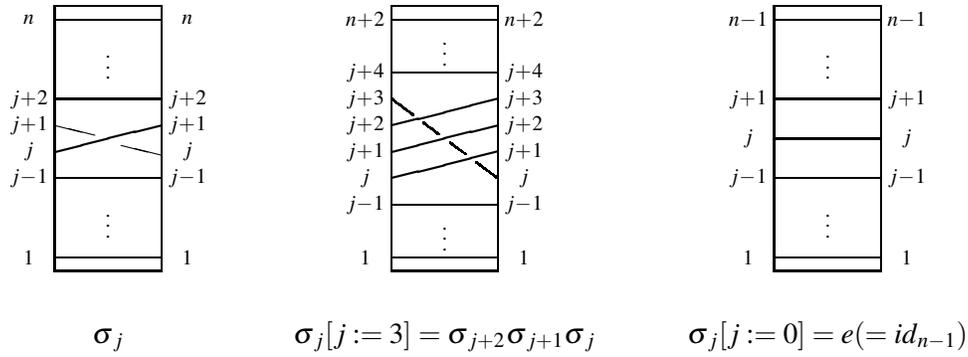

\paragraph{Defining substitutions}
Define the substitution map $(-)[i:=m]:B_n\rightarrow B_{n+m-1}$
for $1\leq i\leq n$ and $m\geq0$ as follows.
\begin{itemize}
\item $e[i:=m]\equiv e$.
\item
$(\ssigma_j s)[i:=m]\equiv\ssigma_{j+m-1}(s[i:=m])$ when
$i\leq j\!-\!1$.
\item
$(\ssigma_j s)[i:=m]\equiv\ssigma_j(s[i:=m])$ when
$i\geq j\!+\!2$.
\item
$(\ssigma_j s)[j:=m]\equiv
\left\{
\begin{array}{ll}
s[j+1:=0] & m=0\\
\ssigma_{j+m-1}\cdots \ssigma_{j+1}\ssigma_j(s[j+1:=m]) & m\geq 1
\end{array}
\right.$
\item
$(\ssigma_j s)[j+1:=m]\equiv
\left\{
\begin{array}{ll}
s[j:=0] & m=0\\
\ssigma_{j}\ssigma_{j+1}\cdots \ssigma_{j+m-1}(s[j:=m]) & m\geq 1
\end{array}
\right.$
\item Similarly for $\ssigma_j^{-1}s$.
\end{itemize}
The substitution map is well-defined: $s[i:=m]$ does not depend on the choice of 
$g_1,\dots,g_k\in\{\ssigma_1^\pm,\dots,\ssigma_n^\pm\}$ such that $s=g_1\cdots g_k$.
Note that $s[i:=1]\equiv s$ holds for any $s\in B_n$ and $i$.
We give some examples of the substitution map in Figure \ref{fig:substitutionmap}.

In the sequel we identify an element of $B_n$ with a braid with $n$ strands.
For a braided term $[s]M$ with
$$
\infer
{x_{s(1)},x_{s(2)},\dots,x_{s(n)}\vdash [s]M}
{x_1,x_2,\dots,x_n\vdash M & s\in B_n}
$$
and a term $y_1,\dots,y_m\vdash N$, we define the substitution $([s]M)[x_i:=N]$
as
$$
([s]M)[x_i:=N]~\equiv~\big[s[s^{-1}(i):=m]\big](M[x_i:=N])
$$

\subsection{Rewriting and Decidability}

Let $\equiv_\mathit{str}$ be the smallest congruence on braided lambda terms
containing the equational theory of braid groups and structural axioms.
We say that a term $M$ (1-step) $\beta\eta$-reduces to $N$ modulo $\equiv_\mathit{str}$
when there exists $M_1$ such that $M\equiv_\mathit{str}M_1$ and 
$M_1$ reduces to $N$ via a single $\beta\eta$-reduction.

\begin{theorem}
The $\beta\eta$-reduction modulo $\equiv_\mathit{str}$
is strong normalizing, and Church-Rosser modulo
$\equiv_\mathit{str}$.
\end{theorem}
\underline{Proof} (outline)
For a braided lambda term $M$, let $u(M)$ be the linear lambda term
obtained by deleting all braids in $M$; thus $u(x)=x$, $u(\lambda x.M)=\lambda x.u(M)$, $u(M\,N)=u(M)\,u(N)$ and $u([s]M)=u(M)$.
Then we have that
\begin{itemize}
\item If $M$ 1-step $\beta\eta$-reduces to $N$, then $u(M)$ 
1-step $\beta\eta$-reduces to $u(N)$ in the linear lambda calculus.
\item If $M\equiv_\mathit{str} M_1$, then $u(M)\equiv u(M_1)$.
\end{itemize}
The strong normalization follows immediately from these observations
as an infinite $\beta\eta$-reduction sequence modulo $\equiv_\mathit{str}$
would give rise to an infnite $\beta\eta$-reduction sequence in the
linear lambda calculus (which of course is strongly normalizing).

The confluence of $\beta$-reduction modulo $\equiv_\mathit{str}$
follows from the observation that when $u(M)$ $\beta$-reduces to 
$N$ in the linear lambda calculus, there exists a braided term $N_0$
such that $M$ $\beta$-reduces to $N_0$ modulo $\equiv_\mathit{str}$
and $u(N_0)=N$ holds.  This does not hold for the $\eta$-reduction:
for 
$M=\lambda y.
\left[\,
\unitlength=.25mm
\begin{picture}(50,10)(-5,0)
\thicklines
\qbezier(0,0)(5,0)(10,5)
\qbezier(10,5)(15,10)(20,10)
\qbezier(0,10)(5,10)(7,8)
\qbezier(13,2)(15,0)(20,0)
\qbezier(20,0)(25,0)(30,5)
\qbezier(30,5)(35,10)(40,10)
\qbezier(20,10)(25,10)(27,8)
\qbezier(33,2)(35,0)(40,0)
\put(-5,10){\makebox(0,0){\footnotesize$y$}}
\put(-5,0){\makebox(0,0){\footnotesize$x$}}

\put(45,10){\makebox(0,0){\footnotesize$y$}}
\put(45,0){\makebox(0,0){\footnotesize$x$}}

\end{picture}
\,\right]
(x\,y)$,
observe that
$u(M)=\lambda y.x\,y$ $\eta$-reduces to $x$
while $M$ is $\eta$-normal in the braided calculus.
Fortunately the $\eta$-postponement holds in this setting
and we obtain the confluence of $\beta\eta$-reduction modulo $\equiv_\mathit{str}$.

\mbox{}

Note that a normal form of $\beta$-reduction modulo $\equiv_\mathit{str}$
is just a $\beta$-normal linear lambda term decorated by braids,
and a normal form of a braided term can be easily obtained by tracing the
normalization of the corresponding linear lambda term. 
Since the word problem for braid groups is decidable \cite{Art25,KT08}
and so is the equational theory of structural axioms, we conclude:
\begin{theorem}
The $\beta\eta$-theory of the braided lambda calculus 
(as given in Section \ref{subsec:eqational-theory}) 
is decidable.
\end{theorem}

\section{Combinatory Logic}

\subsection{Representing Braids by $\bfC^\pm$}

For a braid $s$ with $n$ strands,
let $\lceil s\rceil$
be the combinator 
$$
\lambda f x_{s(1)}\dots x_{s(n)}.[\id_1\otimes s](f\,x_1\,\dots\,x_n)
$$
In particular, when $n=2$ 
$
\lceil\ssigma_1\rceil
=
\left\lceil
\begin{picture}(24,10)(-2,2)
\thicklines
\qbezier(0,0)(5,0)(10,5)
\qbezier(10,5)(15,10)(20,10)
\qbezier(0,10)(5,10)(7,8)
\qbezier(13,2)(15,0)(20,0)
\end{picture}
\right\rceil
=
\bfC^+
$
and
$
\lceil\ssigma_1^{-1}\rceil
=
\left\lceil
\begin{picture}(24,10)(-2,2)
\thicklines
\qbezier(0,10)(5,10)(10,5)
\qbezier(10,5)(15,0)(20,0)
\qbezier(0,0)(5,0)(7,2)
\qbezier(13,8)(15,10)(20,10)
\end{picture}
\right\rceil
=
\bfC^-
$.\\
As usual, we have the combinators $\II\equiv\lambda x.x$ and $\BB\equiv\lambda xyz.x\,(y\,z)$.

\begin{lemma}\label{lem:braids2combinators}
\begin{enumerate}
\item
$\lceil \id_n\rceil~=_{\beta\eta}~\mathbf{I}$.
\item
$\lceil ss'\rceil~=_{\beta\eta}~\mathbf{B}\,\lceil s\rceil\,\lceil s'\rceil$.
\item
$\lceil \id_1\otimes s\rceil ~=_{\beta\eta}~\mathbf{B}\,\lceil s\rceil$.
\item
$\lceil s\otimes\id_1\rceil ~=_{\beta\eta}~\lceil s\rceil$.
\end{enumerate}
\end{lemma}
Below let us write $M^{n+1}\,N$ for$M\,(M^n\,N)$ and $M^0\,N$ for $N$. 
\begin{proposition}\label{prop:braids2combinators}
$\lceil\sigma_i\rceil =_{\beta\eta} \BB^{i-1}\bfC^+$ and
$\lceil\sigma_i^{-1}\rceil=_{\beta\eta}\BB^{i-1}\bfC^-$.
\end{proposition}
Since any braid is given by composing  $e$, $\sigma_i$ and $\sigma_i^{-1}$,
we conclude:
\comment{
$\begin{picture}(24,10)(-2,2)
\thicklines
\qbezier(0,0)(5,0)(10,5)
\qbezier(10,5)(15,10)(20,10)
\qbezier(0,10)(5,10)(7,8)
\qbezier(13,2)(15,0)(20,0)
\end{picture}
$
and
$
\begin{picture}(24,10)(-2,2)
\thicklines
\qbezier(0,10)(5,10)(10,5)
\qbezier(10,5)(15,0)(20,0)
\qbezier(0,0)(5,0)(7,2)
\qbezier(13,8)(15,10)(20,10)
\end{picture}
$}
\begin{theorem}
For any braid $s$, 
$\lceil s\rceil$ is $\beta\eta$-equal to a combinator generated by 
$\mathbf{B}$, $\mathbf{I}$, $\bfC^+$ and $\bfC^-$.
\end{theorem}

\subsection{Combinatory Completeness of $\BB\bfC^\pm\II$}
For the braided term $x_{s(1)},x_{s(2)},\dots,x_{s(n)}\vdash [s]M$, we have 
$$
\begin{array}{rcl}
[s]M
&=_{\beta\eta}&
\lceil s\rceil\,(\lambda x_1\dots x_n.M)\,x_{s(1)}\dots x_{s(n)}\\
\end{array}
$$
because
$$
\begin{array}{rcl}
\lceil s\rceil\,(\lambda x_1\dots x_n.M)\,x_{s(1)}\dots x_{s(n)}
&=&
(\lambda f x_{s(1)}\dots x_{s(n)}.[\id_1\otimes s](f\,x_1\,\dots\,x_n))\,
(\lambda x_1\dots x_n.M)\,x_{s(1)}\dots x_{s(n)}\\
&=&
(\lambda x_{s(1)}\dots x_{s(n)}.[s]((\lambda x_1\dots x_n.M)\,x_1\,\dots\,x_n))\,x_{s(1)}\dots x_{s(n)}\\
&=&
[s]((\lambda x_1\dots x_n.M)\,x_1\,\dots\,x_n)\\
&=&
[s]M
\end{array}
$$
Thus any braided lambda term is equal to a planar lambda term 
(a term which does not involve the braid rule)
enriched with
$\bfC^+$ and $\bfC^-$. In particular, for combinators we have

\begin{theorem}
Any closed term of the braided lambda calculus is $\beta\eta$-equal to a combinator
generated by 
$\mathbf{B}$, $\mathbf{I}$, $\bfC^+$ and $\bfC^-$.
\end{theorem}
This, in the context of combinatory logic, can be thought as a
{\em combinatory completeness}.
Indeed, we have the following translation $(-)^\flat$ from the braided 
lambda calculus to $\BB\bfC^\pm\II$-terms.
$$
\begin{array}{rclrclrcl}
x^\flat &\equiv& x &~~~
(M\,N)^\flat &\equiv& M^\flat\,N^\flat ~~~&
(\lambda x.M)^\flat &\equiv& \lambda^*x.M^\flat\\
\multicolumn{9}{c}
{([s]M)^\flat ~\equiv~
\lceil s\rceil\,(\lambda^* x_1\dots x_n.M^\flat)\,x_{s(1)}\dots x_{s(n)}
~~~~(\mathrm{cxt}(M)=x_1,\dots,x_n)}\\
\multicolumn{4}{c}
{\lambda^* x.x ~\equiv~ \II} 
& 
\multicolumn{5}{c}
{\lambda^* x.P\,Q ~\equiv~
\left\{
\begin{array}{ll}
\bfC^+\,(\lambda^*x.P)\,Q & (x\in\mathrm{fv}(P))\\
\BB\,P\,(\lambda^*x.Q) & (x\in\mathrm{fv}(Q))
\end{array}
\right.
}\\
\end{array}
$$
\comment{
$$
\begin{array}{rcl}
x^\flat &\equiv& x\\
(M\,N)^\flat &\equiv& M^\flat\,N^\flat\\
([s]M)^\flat &\equiv& 
\lceil s\rceil\,(\lambda^* x_1\dots x_n.M^\flat)\,x_{s(1)}\dots x_{s(n)}
~~~~(\mathrm{cxt}(M)=x_1,\dots,x_n)\\
(\lambda x.M)^\flat &\equiv& \lambda^*x.M^\flat\\
\lambda^* x.x &\equiv& \II\\
\lambda^* x.P\,Q &\equiv& 
\left\{
\begin{array}{ll}
\bfC^+\,(\lambda^*x.P)\,Q & (x\in\mathrm{fv}(P))\\
\BB\,P\,(\lambda^*x.Q) & (x\in\mathrm{fv}(Q))
\end{array}
\right.\\
\end{array}
$$}%
(To be precise, this determines a translation on terms modulo $\beta\eta$-equality,
because 
Lemma \ref{lem:braids2combinators}
and Proposition \ref{prop:braids2combinators} define
$\lceil s\rceil$ only up to $\beta\eta$-equality. 
For instance, $e=\sigma_1\sigma_1^{-1}$ in $B_2$
and 
$\lceil e\rceil=\lambda fxy.fxy$ while $\lceil \sigma_1\sigma_1^{-1}\rceil=\BB\mathbf{C}^+\mathbf{C}^-$, and they are $\beta\eta$-equal.)

\begin{example}
As an example involving a fairly complex braid, 
let us consider a {Celtic C-combinator} (inspired by the traditional
Celtic braid):

$$
\lambda fxyz.\left[(\sigma_2\sigma_1^{-1}\sigma_3^{-1})^4\sigma_2\right](f\,y\,x\,z)
~~=~~
{\lambda f x y z}.
\left[
\setlength\unitlength{1pt}
\begin{picture}(140,30)(-20,20)
\thicklines
\put(-10,40){\line(1,0){10}}%
\qbezier(-10,27)(-8,27)(-2,22)%
\qbezier(-10,13)(-8,13)(0,20)%
\put(-10,0){\line(1,0){10}}%
\celticbraidcolor{0}{0}{}{}{}{}%
\celticbraidcolor{25}{0}{}{}{}{}
\celticbraidcolor{50}{0}{}{}{}{}
\celticbraidcolor{75}{0}{}{}{}{}
\put(100,40){\line(1,0){10}}%
\qbezier(100,20)(108,27)(110,27)%
\qbezier(102,18)(108,13)(110,13)%
\put(100,00){\line(1,0){10}}%
\put(-15,0){\makebox(0,0){$f$}}
\put(-15,13){\makebox(0,0){$x$}}
\put(-15,26){\makebox(0,0){$y$}}
\put(-15,40){\makebox(0,0){$z$}}
\put(115,0){\makebox(0,0){$f$}}
\put(115,13){\makebox(0,0){$y$}}
\put(115,26){\makebox(0,0){$x$}}
\put(115,40){\makebox(0,0){$z$}}
\end{picture}
\right]
({ f\, y\, x\, z})
$$%
Thanks to the combinatory completeness and the translation above,
we have that
this combinator is $\beta\eta$-equal to
$$
\begin{array}{l}
\BB\,\Cp\,(\Cm\,(\BB\,(\BB\,\Cm))\,
\\
~~~~
(\BB\,\Cp\,(\Cm\,(\BB\,(\BB\,\Cm))\,
\\
~~~~~~~~
(\BB\,\Cp\,(\Cm\,(\BB\,(\BB\,\Cm))\,
\\
~~~~~~~~~~~~
(\BB\,\Cp\,(\Cm\,(\BB\,(\BB\,\Cm))\,\Cp)))))))
\end{array}
$$
built from $\mathbf{B}$, $\mathbf{I}$, $\bfC^+$ and $\bfC^-$.
\end{example}

Therefore it is possible to formulate a {\em braided combinatory logic}
with constants $\BB$, $\mathbf{C}^\pm$, $\II$ and an appropriate set of
axioms (say $\mathcal{A}$) ensuring (i) $M=_\mathcal{A}M'$ implies 
$\lambda^*x.M=_\mathcal{A}\lambda^*x.M'$
and (ii) $s=s'$ in $B_n$ implies $\lceil s\rceil=_\mathcal{A}\lceil s'\rceil$.
Finding a complete (hopefully finite) axiomatization (which should 
satisfy (i) and (ii) above)
is left as future work.

For comparison, in Figure \ref{fig:BCI} we give an axiomatization of the linear combinatory logic $\BB\CC\II$
which is sound and complete for the $\beta\eta$-theory of the linear lambda calculus.\footnote{%
This axiomatization is our own version (and might contain some redundancies); 
we were unable to find such a complete axiomatization
of $\BB\CC\II$ in the literature, though we think that an axiomatization like ours
should be known to specialists.
For reference, we include an outline of the proof of completeness in Appendix \ref{sec:BCI}.
}
We expect that a complete axiomatization of $\BB\CC^\pm\II$
can be given like this axiomatization of $\BB\CC\II$,
 with some needed modifications.
For instance, {\em Reidemeister II} and {\em Reidemeister III}
should be replaced by the braided versions
$$
\begin{array}{rcll}
\BB\,\bfC^\pm\,\bfC^\mp &=& \II &\mbox{\em Reidemeister II}\\
\BB\,(\BB\,\bfC^+)\,(\BB\,\bfC^+\,(\BB\,\bfC^+))
&=&
\BB\,\bfC^+\,(\BB\,(\BB\,\bfC^+)\,\bfC^+)
& \mbox{\em Reidemeister III}\\
\end{array}
$$
which amount to $\lceil\sigma_1\sigma_1^{-1}\rceil=\lceil\sigma_1^{-1}\sigma_1\rceil=\II$
and $\lceil\sigma_1\sigma_2\sigma_1\rceil=\lceil\sigma_2\sigma_1\sigma_2\rceil$.
It seems much more difficult  to find a braided variant of the axiom (C):
when the terms $L$, $M$, $N$ have $l$, $m$, $n$ free variables respectively,
we have
$$
\CC^+\,L\,M\,N~=~
\left[
\begin{picture}(60,35)(0,28)\thicklines
\qbezier(10,55)(10,55)(28,46)\qbezier(42,39)(50,35)(50,35)
\qbezier(10,45)(10,45)(18,41)\qbezier(32,34)(50,25)(50,25)
\put(10,35){\line(2,1){40}}
\put(10,25){\line(2,1){40}}
\idline{10}{15}{40}
\idline{10}{5}{40}
\put(8,50){\makebox(0,0){\{ }} \put(2,50){\makebox(0,0){\small$n$}}
\put(8,30){\makebox(0,0){\{ }} \put(2,30){\makebox(0,0){\small$m$}}
\put(8,10){\makebox(0,0){\{ }} \put(2,10){\makebox(0,0){\small$l$}}
\put(30,42){\makebox(0,0){$\cdot$}}
\put(30,40){\makebox(0,0){$\cdot$}}
\put(30,38){\makebox(0,0){$\cdot$}}
\put(30,12){\makebox(0,0){$\cdot$}}
\put(30,10){\makebox(0,0){$\cdot$}}
\put(30,8){\makebox(0,0){$\cdot$}}
\end{picture}
\right]
(L\,N\,M)
$$
where the braid in the right hand side of the equation is not an identity unless
$m$ or $n$ is zero, and the corresponding $\BB\CC^\pm\II$-term 
contains $mn$ $\CC^+$s. This suggests that finding a finite axiomatization
is not an obvious task.

\begin{figure}
$$
\begin{array}{rcll}
\BB\,L\,M\,N &=& L\,(M\,N) & \mathrm{(B)}\\
\CC\,L\,M\,N &=& L\,N\,M & \mathrm{(C)}\\
\II\,M &=& M & \mathrm{(I)}\\
\BB\,\II &=& \II \\
\CC\,\BB\,\II &=& \II \\
\BB\,(\BB\,\BB)\,\BB &=& \BB\,(\CC\,\BB\,\BB)\,(\BB\,\BB\,\BB) \\
\BB\,(\BB\,\CC)\,(\BB\,\BB\,\BB) &=& \BB\,(\CC\,\BB\,\CC)\,(\BB\,\BB\,\BB)\\
\BB\,(\BB\,\BB)\,\CC &=& \BB\,\CC\,(\BB\,(\BB\,\CC)\,\BB)\\
\BB\,\CC\,\CC &=& \II & \mbox{\em Reidemeister II}\\
\BB\,(\BB\,\CC)\,(\BB\,\CC\,(\BB\,\CC)) &=& 
\BB\,\CC\,(\BB\,(\BB\,\CC)\,\CC) & \mbox{\em Reidemeister III}\\
\end{array}
$$
\caption{A complete axiomatization of $\BB\CC\II$}
\label{fig:BCI}
\end{figure}

\section{Semantics}

\subsection{Categorical Models}
A model of the braided lambda calculus (without $\eta$) can be given by an object $X$
in a braided monoidal closed category \cite{JS93}
such that the internal hom $[X,X]$ 
is a retract of $X$. 
The situation is largely the same as that of the models of the untyped lambda calculus
given by reflexive objects in cartesian closed categories. Let us sketch how it works. 
Let $\mathit{ev}:[X,X]\otimes X\rightarrow X$ be the evaluation map given by
the monoidal closure, and write $\Lambda(f):\Gamma\rightarrow [X,X]$ for
the currying of an arrow $f:\Gamma\otimes X\rightarrow X$.
Assume an arrow $\varphi:[X,X]\rightarrow X$ with a right inverse
$\psi:X\rightarrow [X,X]$. Then we can interpret
a braided lambda term $x_1,\dots,x_n\vdash M$ as an arrow
$\sem{x_1,\dots,x_n\vdash M}$ from 
$X^{\otimes n}=\overbrace{X\otimes\dots\otimes X}^n$ to $X$
as follows.
$$
\begin{array}{rcl}
\sem{x\vdash x} &=&\mathit{id}_X\\
\sem{\Gamma\vdash \lambda x.M}&=&
\Lambda(\sem{\Gamma,x\vdash M});\varphi\\
\sem{\Gamma,\Delta\vdash M\,N}&=&
(\sem{\Gamma\vdash M};\psi\otimes\sem{\Delta\vdash N});\mathit{ev}\\
\sem{x_{s(1)},\dots,x_{s(n)}\vdash [s]M}&=&\sem{s};\sem{x_1,\dots,x_n\vdash M}
\end{array}
$$
where $;$ denotes the relational composition, and the interpretation $\sem{s}$
is the interpretation of the  braid $s$ on $X^{\otimes n}$.
The $\beta$-equality is validated because $\varphi;\psi=\mathit{id}_{[X,X]}$
hold.
An extensional model (i.e., validating $\eta$) is given by an $X$ such that
$[X,X]$ is isomorphic to $X$, i.e., $\psi=\varphi^{-1}$.

There are plenty of braided monoidal closed categories in the literature ---
many of them are found in the context of {representation theory of 
quantum groups} \cite{Tur94}. However, finding a braided monoidal closed category with
a non-trivial reflexive object is not easy --- impossible if we stick to 
finite dimensional linear representations, as the dimension of $[X,X]$ 
is strictly higher than that of $X$ unless $X$ is one-dimensional. 
Below we present models using braided relational semantics \cite{Has12}
where the problem of dimensions disappears.

\subsection{A Crossed $G$-Set Model of Finite Binary Trees}

Fix a group $G=(G,e,\cdot,(-)^{-1})$. 
Recall that a {\em crossed $G$-set} \cite{Whi49}
is a set $X$ equipped with a $G$-action
$\bullet:G\times X\rightarrow X$ and a valuation map $|\_|:X\rightarrow G$
satisfying $|g\bullet x|=g|x|g^{-1}$ for $g\in G$ and $x\in X$. 
There is a {\em ribbon category} \cite{Shu94,Tur94} $\mathbf{XRel}(G)$
 whose objects are crossed $G$-sets and
a morphism from $(X,\bullet,|\_|)$ to $(Y,\bullet,|\_|)$
is a binary relation $r\subseteq X\times Y$ between $X$ and $Y$
such that $(x,y)\in r$ implies $|x|=|y|$ as well as $(g\bullet x,g\bullet y)\in r$
for any $g\in G$ \cite{Has12}. 
The dual of a crossed $G$-set $X=(X,\bullet,|\_|)$ is $X^*=(X,\bullet,|\_|^{-1})$. 
The tensor of $X=(X,\bullet,|\_|)$ and $Y=(Y,\bullet,|\_|)$ is
$X\otimes Y=(X\times Y,(g,(x,y))\mapsto (g\bullet x,g\bullet y),(x,y)\mapsto |x||y|)$.
For this monoidal structure we have a braiding
$\sigma_{X,Y}:X\otimes Y\stackrel{\cong}{\rightarrow} Y\otimes X$ 
as
$$\sigma_{X,Y}=\{((x,y),(|x|\bullet y,x))~|~x\in X,y\in Y\}.$$
See \cite{Has12} for further details  of $\mathbf{XRel}(G)$.

Below we will give a crossed $G$-set
$\mathcal{T}$ such that the internal hom 
$[\mathcal{T},\mathcal{T}]=\mathcal{T}\otimes\mathcal{T}^*$
is a retract of $\mathcal{T}$, which forms a model of the braided lambda calculus.

Let $\mathcal{T}$ be the set of binary trees 
whose leaves are labelled by elements of $G$
(or the implicational formulas generated from $G$):
$$t~::=~g~|~t\rarrow t ~~~~~ (g\in G)$$
$\mathcal{T}$ is a crossed $G$-set with 
the valuation $|\_|:\mathcal{T}\rightarrow G$ given by
$|g|=g$ and $|x\rarrow y|=|x||y|^{-1}$
and the $G$-action $\bullet:G\times\mathcal{T}\rightarrow\mathcal{T}$
given by 
$$
g\bullet h=ghg^{-1}~~(h\in G),~~~~
g\bullet(x\rarrow y)=(g\bullet x)\rarrow(g\bullet y)
$$
Moreover the map $\varphi:\mathcal{T}\times\mathcal{T}\rightarrow\mathcal{T}$
sending $(x,y)$ to $x\rarrow y$ gives a morphism  
$$\varphi=\{((x,y),x\rarrow y)~|~x,y\in\mathcal{T}\}:
\mathcal{T}\otimes\mathcal{T}^*\rightarrow\mathcal{T}$$ 
in $\mathbf{XRel}(G)$, with a right inverse
$\psi=\{(x\rarrow y,(x,y))~|~x,y\in\mathcal{T}\}:
\mathcal{T}\rightarrow\mathcal{T}\otimes\mathcal{T}^*$.
It follows that we can model the untyped braided lambda calculus
(without $\eta$) 
using $\mathcal{T}$ as follows. A term $x_1,\dots,x_n\vdash M$ is 
interpreted as a relation $r$ from $\mathcal{T}^n$ to $\mathcal{T}$
such that $((u_1,\dots,u_n),a)\in r$ implies $|u_1|\cdots|u_n|=|a|$ 
as well as $((g\bullet u_1,\dots,g\bullet u_n),g\bullet a)\in r$ for any $g\in G$.
In particular, a closed term is interpreted as a subset of $\{x\in\mathcal{T}~|~|x|=e\}$
closed under the $G$-action.
$$
\begin{array}{rcl}
\sem{x\vdash x} &=&\{(a,a)~|~a\in\mathcal{T}\}\\
\sem{\Gamma\vdash \lambda x.M}&=&
\{(\vec{u},b\rarrow a)~|~((\vec{u},a),b)\in\sem{\Gamma,x\vdash M}\}\\
\sem{\Gamma,\Delta\vdash M\,N}&=&
\{((\vec{u},\vec{v}),b)~|~\exists a~(\vec{u},b\rarrow a)\in\sem{\Gamma\vdash M}~\&~(\vec{v},a)\in\sem{\Delta\vdash N}\}\\
\sem{x_{s(1)},\dots,x_{s(n)}\vdash [s]M}&=&\sem{s};\sem{x_1,\dots,x_n\vdash M}
\end{array}
$$
where the interpretation $\sem{s}$ of a braid $s$ is built from 
$$
\begin{array}{rcl}
\sem{
\begin{picture}(24,10)(-2,2)
\thicklines
\qbezier(0,0)(5,0)(10,5)
\qbezier(10,5)(15,10)(20,10)
\qbezier(0,10)(5,10)(7,8)
\qbezier(13,2)(15,0)(20,0)
\end{picture}
}
&=&
\{((a,b),(|a|\bullet b, a))~|~a,b\in\mathcal{T}\}
\\
\sem{
\begin{picture}(24,10)(-2,2)
\thicklines
\qbezier(0,10)(5,10)(10,5)
\qbezier(10,5)(15,0)(20,0)
\qbezier(0,0)(5,0)(7,2)
\qbezier(13,8)(15,10)(20,10)
\end{picture}
}
&=&
\{((a,b),(b, |b|^{-1}\bullet a))~|~a,b\in\mathcal{T}\}
\end{array}
$$
For instance, the braided $\mathbf{C}$ combinators are interpreted as 
$$
\begin{array}{rcl}
\sem{\mathbf{C}^+}
&=&
\{((z\rarrow |x|\bullet y)\rarrow x)\rarrow ((z\rarrow x)\rarrow y)
   ~|~x,y,z\in\mathcal{T}\}
\\
\sem{\mathbf{C}^-}
&=&
\{((z\rarrow y)\rarrow |y|^{-1}\bullet x)\rarrow ((z\rarrow x)\rarrow  y)
   ~|~x,y,z\in\mathcal{T}\}
\end{array}
$$
This model does not validate the $\eta$-equality:
$$
\sem{x\vdash \lambda y.x\,y}=\{(b\rarrow a,b\rarrow a)~|~a,b\in\mathcal{T}\}
\not=
\sem{x\vdash x}.
$$
This is because $\varphi$ is not an isomorphism; the right inverse $\psi$ 
cannot map leaves of $\mathcal{T}$ to elements of $\mathcal{T}\otimes\mathcal{T}^*$.
(It might be tempting to remedy this by taking a quotient of $\mathcal{T}$ by
identifying $x\rarrow e$ with $x$, as suggested by an anonymous  reviewer.
This certainly makes $\psi;\varphi=\mathit{id}_X$ and the $\eta$-equality becomes
valid. Unfortunately, on this quotient, $\varphi;\psi$ is no longer the identity, and the $\beta$-equality
becomes invalid.)

\subsection{An Extensional Crossed $G$-Set Model of Infinite Binary Trees}

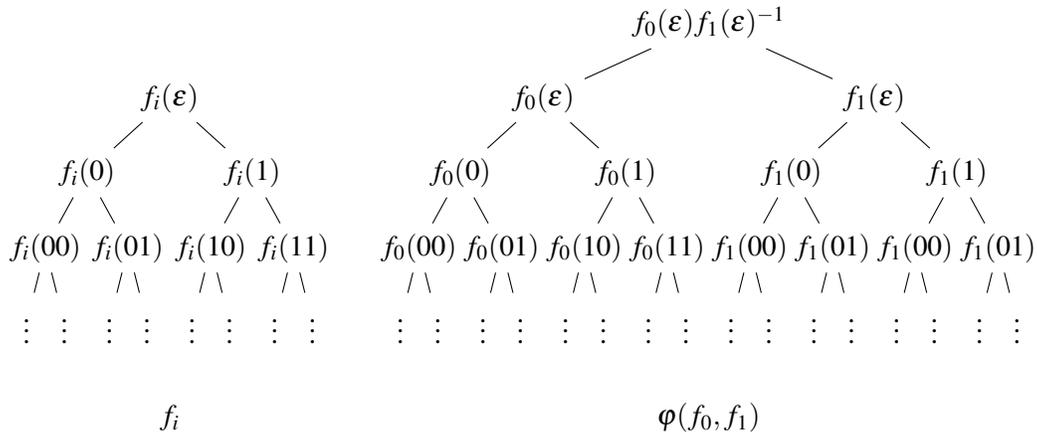
\begin{figure}
\begin{center}
\begin{tabular}{cc}
\begin{tikzpicture}[level distance=1cm,
  level 1/.style={sibling distance=2.2cm},
  level 2/.style={sibling distance=1.1cm},
  level 3/.style={sibling distance=.5cm}]
  \node {$f_i(\epsilon)$}
    child {node {$f_i(0)$}
      child {node {$f_i(00)$} child {node {$\vdots$}} child {node {$\vdots$}}}
      child {node {$f_i(01)$} child {node {$\vdots$}} child {node {$\vdots$}}}
    }
    child {node {$f_i(1)$}
    child {node {$f_i(10)$} child {node {$\vdots$}} child {node {$\vdots$}}}
      child {node {$f_i(11)$} child {node {$\vdots$}} child {node {$\vdots$}}}
    };
\end{tikzpicture}
&
\begin{tikzpicture}[level distance=1cm,
  level 1/.style={sibling distance=4.4cm},
  level 2/.style={sibling distance=2.2cm},
  level 3/.style={sibling distance=1.1cm},
  level 4/.style={sibling distance=.5cm}]
  \node {$f_0(\epsilon)f_1(\epsilon)^{-1}$}
    child {node {$f_0(\epsilon)$}
      child {node {$f_0(0)$} child {node {$f_0(00)$} child {node {$\vdots$}} child {node {$\vdots$}}} child {node {$f_0(01)$} child {node {$\vdots$}} child {node {$\vdots$}}}}
      child {node {$f_0(1)$} child {node {$f_0(10)$} child {node {$\vdots$}} child {node {$\vdots$}}} child {node {$f_0(11)$} child {node {$\vdots$}} child {node {$\vdots$}}}}
    }
    child {node {$f_1(\epsilon)$}
    child {node {$f_1(0)$} child {node {$f_1(00)$} child {node {$\vdots$}} child {node {$\vdots$}}} child {node {$f_1(01)$} child {node {$\vdots$}} child {node {$\vdots$}}}}
      child {node {$f_1(1)$} child {node {$f_1(00)$} child {node {$\vdots$}} child {node {$\vdots$}}} child {node {$f_1(01)$} child {node {$\vdots$}} child {node {$\vdots$}}}}
    };
\end{tikzpicture}
\\
\\
$f_i$ & $\varphi(f_0,f_1)$
\end{tabular}
\end{center}
\comment{%
\begin{center}
\begin{tabular}{cc}
\begin{tikzpicture}[level distance=1cm,
  level 1/.style={sibling distance=2.2cm},
  level 2/.style={sibling distance=1.1cm},
  level 3/.style={sibling distance=.5cm}]
  \node {$f_i(\epsilon)$}
    child {node {$f_i(0)$}
      child {node {$f_i(00)$} child  child}
      child {node {$f_i(01)$} child  child}
    }
    child {node {$f_i(1)$}
    child {node {$f_i(10)$} child child}
      child {node {$f_i(11)$} child child}
    };
\end{tikzpicture}
&
\begin{tikzpicture}[level distance=1cm,
  level 1/.style={sibling distance=4.4cm},
  level 2/.style={sibling distance=2.2cm},
  level 3/.style={sibling distance=1.1cm},
  level 4/.style={sibling distance=.5cm}]
  \node {$f_0(\epsilon)f_1(\epsilon)^{-1}$}
    child {node {$f_0(\epsilon)$}
      child {node {$f_0(0)$} child {node {$f_0(00)$} child child} child {node {$f_0(01)$} child child}}
      child {node {$f_0(1)$} child {node {$f_0(10)$} child child} child {node {$f_0(11)$} child child}}
    }
    child {node {$f_1(\epsilon)$}
    child {node {$f_1(0)$} child {node {$f_1(00)$} child child} child {node {$f_1(01)$} child child}}
      child {node {$f_1(1)$} child {node {$f_1(00)$} child child} child {node {$f_1(01)$} child child}}
    };
\end{tikzpicture}
\\
\\
$f_i$ & $\varphi(f_0,f_1)$
\end{tabular}
\end{center}
}%
\caption{$\varphi(f_0,f_1)$}
\label{fig:verphi}
\end{figure}

Now we expand $\mathcal{T}$ to a crossed $G$-set of infinite binary trees.
Let 
$$
\mathcal{D}=\{f:\{0,1\}^*\rightarrow G~|~f(w)=f(w0)\cdot f(w1)^{-1}\}
$$
$\mathcal{D}$ is a crossed $G$-set with $|f|=f(\epsilon)$ and
$(g\bullet f)(w)=g\cdot f(w)\cdot g^{-1}$.
Its dual $\mathcal{D}^*$ is identical to $\mathcal{D}$ except the
valuation $|f|=f(\epsilon)^{-1}$.
There is an isomorphism $\varphi:\mathcal{D}\otimes\mathcal{D}^*\stackrel{\simeq}{\rightarrow}\mathcal{D}$ induced by the
bijective map $\varphi:\mathcal{D}^2\rightarrow\mathcal{D}$
given by (see Figure \ref{fig:verphi})
$$
\left\{
\begin{array}{lcl}
\varphi(f_0,f_1)(\epsilon)&=&f_0(\epsilon)f_1(\epsilon)^{-1}\\
\varphi(f_0,f_1)(0w)&=&f_0(w)\\
\varphi(f_0,f_1)(1w)&=&f_1(w)\\
\end{array}
\right.
$$
Note that $\varphi^{-1}(f)=(\lambda w.f(0w),\lambda w.f(1w))$ holds.
Also  $\mathcal{D}\cong\mathcal{D}^*$ with
$f\mapsto f^*=\varphi(\lambda w.f(1w),\lambda w.f(0w))$
(thus $f^*(\epsilon)=f(\epsilon)^{-1}$, $f^*(0w)=f(1w)$ and $f^*(1w)=f(0w)$).
$\mathcal{D}$ is a model of the braided lambda calculus validating the
$\eta$ equality. The interpretation of terms is essentially the same as the
case of $\mathcal{T}$, with $x \rarrow y$ replaced by $\varphi(x,y)$.

\comment{
\begin{itemize}
\item $\sem{x\vdash x}=\{(a,a)~|~a\in\mathcal{D}\}$
\item $\sem{\Gamma\vdash \lambda x.M}=
\{(\vec{u},\varphi(b,a))~|~((\vec{u},a),b)\in\sem{\Gamma,x\vdash M}\}$
\item $\sem{\Gamma,\Delta\vdash M\,N}=
\{((\vec{u},\vec{v}),b)~|~\exists a~(\vec{u},\varphi(b,a))\in\sem{\Gamma\vdash M}\&(\vec{v},a)\in\sem{\Delta\vdash N}\}$
\item $\sem{\Gamma'\vdash [s]M}=\sem{s};\sem{\Gamma\vdash M}$
\end{itemize}
$$
\begin{array}{rcl}
\sem{
\begin{picture}(24,10)(-2,2)
\thicklines
\qbezier(0,0)(5,0)(10,5)
\qbezier(10,5)(15,10)(20,10)
\qbezier(0,10)(5,10)(7,8)
\qbezier(13,2)(15,0)(20,0)
\end{picture}
}
&=&
\{((a,b),(|a|\bullet b, a))~|~a,b\in\mathcal{D}\}
\\
\sem{
\begin{picture}(24,10)(-2,2)
\thicklines
\qbezier(0,10)(5,10)(10,5)
\qbezier(10,5)(15,0)(20,0)
\qbezier(0,0)(5,0)(7,2)
\qbezier(13,8)(15,10)(20,10)
\end{picture}
}
&=&
\{((a,b),(b, |b|^{-1}\bullet a))~|~a,b\in\mathcal{D}\}
\end{array}
$$
}

\begin{remark}[a two-objects ribbon category, and the tangled lambda calculus]
Since $\mathcal{D}\cong\mathcal{D}^*\cong\mathcal{D}\otimes\mathcal{D}$,
the full subcategory of $\mathbf{XRel}(G)$ with just   $\mathcal{D}$
and the tensor unit $I$
is a ribbon category. This also means that, with $\mathcal{D}$, we can interpret not just braids but also
{framed tangles} (ribbons). Thus $\mathcal{D}$ is a model of a
``{tangled lambda calculus}'' in which we should be able to express a term involving tangles like
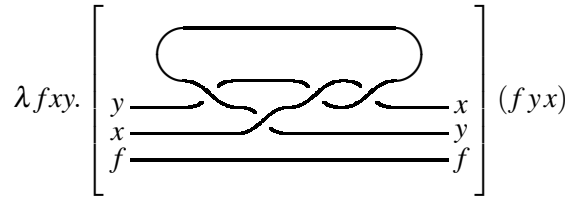
\comment{%
$$
\lambda fxy.
\left[
\unitlength=.2mm
\begin{picture}(140,50)(-10,15)
\thicklines
\trace{20}{50}{80}{10}
\braidInv{20}{20}{20}
\braid{40}{0}{20}
\braid{60}{20}{20}
\braid{80}{20}{20}
\idline{0}{-20}{120}
\idline{0}{0}{40}
\idline{60}{0}{60}
\idline{0}{20}{20}
\idline{100}{20}{20}
\idline{40}{40}{20}
\put(125,-20){\makebox(0,0){$f$}}
\put(125,0){\makebox(0,0){$y$}}
\put(125,20){\makebox(0,0){$x$}}
\put(-5,-20){\makebox(0,0){$f$}}
\put(-5,0){\makebox(0,0){$x$}}
\put(-5,20){\makebox(0,0){$y$}}
\end{picture}
\right](f\,y\,x)
$$
}%
$$
\lambda fxy.
\left[
\begin{picture}(140,40)(-10,10)
\thicklines
\trace{20}{30}{80}{10}
\put(20,10){\usebox{\braidNN}}
\put(40,0){\usebox{\braidPP}}
\put(60,10){\usebox{\braidPP}}
\put(80,10){\usebox{\braidPP}}
\idline{0}{-10}{120}
\idline{0}{0}{40}
\idline{60}{0}{60}
\idline{0}{10}{20}
\idline{100}{10}{20}
\idline{40}{20}{20}
\put(125,-10){\makebox(0,0){$f$}}
\put(125,0){\makebox(0,0){$y$}}
\put(125,10){\makebox(0,0){$x$}}
\put(-5,-10){\makebox(0,0){$f$}}
\put(-5,0){\makebox(0,0){$x$}}
\put(-5,10){\makebox(0,0){$y$}}
\end{picture}
\right](f\,y\,x)
$$
Such a tangled lambda calculus is yet to be studied;
defining substitution already seems to be much harder than the
braided case. Also it might be more appropriate to use traced monoidal
closed categories \cite{Has09} as semantic models rather than ribbon categories.
\end{remark}

\section{Conclusion}

We introduced the syntax and semantics of an untyped braided lambda calculus.
Future work will include
the typed variants, complete axiomatization of the braided combinatory logic,
extension to the tangled lambda calculus, and applications
to novel computational models making use of braids, most notably topological
quantum computation.

\paragraph{Acknowledgements}
I thank Haruka Tomita for stimulating discussions
related to this work, and the anonymous
reviewers for their helpful comments.
This work was supported by 
JSPS KAKENHI Grant Numbers JP18K11165, JP21K11753
and
JST ERATO Grant Number JPMJER1603, Japan.

\nocite{*}
\bibliographystyle{eptcs}
\bibliography{blambda}

\appendix

\newcommand{\BCI}{\mathbf{BCI}}
\newcommand{\llambda}{\lambda_{\mathit{lin}}}

\section{Axiomatizing BCI}
\label{sec:BCI}

Let $\llambda$ be the set of linear lambda terms 
and $\BCI$ be the set of terms generated by variables (each occurring just once), $\BB$, $\CC$, $\II$ and application.
Let $=_\BCI$ be the smallest congruence on $\BCI$ satisfying the  axioms
in Figure \ref{fig:BCI}.
Define translations $(-)^\sharp:\BCI\rightarrow\llambda$ and $(-)^\flat:\llambda\rightarrow\BCI$
by
$$
\begin{array}{c}
\BB^\sharp ~\equiv~ \lambda xyz.x\,(y\,z)
~~~~~~
\CC^\sharp ~\equiv~ \lambda xyz.x\,z\,y
~~~~~~
\II^\sharp ~\equiv~ \lambda x.x
\\
(P\,Q)^\sharp ~\equiv~ P^\sharp\, Q^\sharp
~~~~~~
x^\sharp ~\equiv~ x
\\
\\
(\lambda x.M)^\flat ~\equiv~ \lambda^* x.M^\flat
~~~~~~
(M\,N)^\flat ~\equiv~ M^\flat\,N^\flat
~~~~~~
x^\flat ~\equiv~ x
\\
\lambda^* x.x ~\equiv~ \II
~~~~~~
\lambda^* x.M\,N ~\equiv~
\left\{
\begin{array}{ll}
\CC\,(\lambda^*x.M)\,N & (x\in\mathrm{fv}(M))\\
\BB\,M\,(\lambda^*x.N) & (x\in\mathrm{fv}(N))
\end{array}
\right.\\
\end{array}
$$
\comment{%
$$
\begin{array}{rcl}
\BB^\sharp &\equiv& \lambda xyz.x\,(y\,z)\\
\CC^\sharp &\equiv& \lambda xyz.x\,z\,y\\
\II^\sharp &\equiv& \lambda x.x\\
(P\,Q)^\sharp &\equiv& P^\sharp\, Q^\sharp\\
x^\sharp &\equiv& x\\
\\
(\lambda x.M)^\flat &\equiv& \lambda^* x.M^\flat\\
(M\,N)^\flat &\equiv& M^\flat\,N^\flat\\
x^\flat &\equiv& x\\
\lambda^* x.x &\equiv& \II\\
\lambda^* x.M\,N &\equiv& 
\left\{
\begin{array}{ll}
\CC\,(\lambda^*x.M)\,N & (x\in\mathrm{fv}(M))\\
\BB\,M\,(\lambda^*x.N) & (x\in\mathrm{fv}(N))
\end{array}
\right.\\
\end{array}
$$
}%
We show that these translations give isomorphisms between
the equational theories.
It is routine to see:
\begin{lemma}\label{lem:eqBCI-implies-eqLLambda}
$P=_\BCI Q$ implies $P^\sharp=_{\beta\eta}Q^\sharp$.
\end{lemma}
The following lemma is crucial and the most difficult:
\begin{lemma}\label{lem:lambda-star-cong}
$P=_\BCI Q$ implies $\lambda^* x.P=_\BCI \lambda^*x.Q$.
\end{lemma}
\underline{Proof}
For each axiom $P=Q$ with free $x$ we show $\lambda^*x.P=\lambda^*x.Q$.
The relevant cases are (B), (C) and (I).
For the case of (I), we are to show $\lambda^*x.\II\,M=\lambda^* x.M$
with free $x$ in $M$,
which follows from
$$\lambda^*x.\II\,M
\equiv
\BB\,\II\,(\lambda^*x.M)
=
\II\,(\lambda^*x.M)
=
\lambda^*x.M
$$
The case of (B) contains three sub-cases depending on where the free $x$ occurs.
For instance, showing $\lambda^*x.\BB\,L\,M\,N=\lambda^*x.L\,(M\,N)$
with free $x$ in $N$ amounts to showing 
$\BB\,L\,(\BB\,M\,(\lambda^*x.N))=\BB\,(\BB\,L\,M)\,(\lambda^*x.N)$
for which it suffices to show the associativity $\BB\,L\,(\BB\,M\,N)=\BB\,(\BB\,L\,M)\,N$.  
$$
\begin{array}{rcll}
\BB\,L\,(\BB\,M\,N)
&=&
\BB\,(\BB\,L)\,(\BB\,M)\,N & (B)\\
&=&
\BB\,\BB\,\BB\,L\,(\BB\,M)\,N & (B) \\
&=&
\BB\,(\BB\,\BB\,\BB\,L)\,\BB\,M\,N & (B)\\
&=&
\CC\,\BB\,\BB\,(\BB\,\BB\,\BB\,L)\,M\,N & (C)\\
&=&
\BB\,(\CC\,\BB\,\BB)\,(\BB\,\BB\,\BB)\,L\,M\,N &(B)\\
&=&
\BB\,(\BB\,\BB)\,\BB\,L\,M\,N & (\BB\,(\BB\,\BB)\,\BB~=~\BB\,(\CC\,\BB\,\BB)\,(\BB\,\BB\,\BB))\\
&=&
\BB\,\BB\,(\BB\,L)\,M\,N & (B)\\
&=&
\BB\,(\BB\,L\,M)\,N & (B)
\end{array}
$$
Other two sub-cases of (B) and three sub-cases of (C) are similar (and more lengthy).

\begin{lemma}\label{lem:eqLLambda-implies-eqBCI}
$M=_{\beta\eta} N$ implies $M^\flat=_\BCI N^\flat$.
\end{lemma}
\underline{Proof}
The most nontrivial part is to show that $M=_{\beta\eta}N$ implies
$(\lambda x.M)^\flat=_\BCI(\lambda x.N)^\flat$, which follows from
Lemma \ref{lem:lambda-star-cong}.

\mbox{}\\
The following two lemmas are fairly straightforward.
\begin{lemma}\label{lem:BCI2LLambda2BCI-is-identity} 
$(P^\sharp)^\flat=_\BCI P$. 
\end{lemma}
\comment{%
\underline{Proof}
It suffices to show $(\BB^\sharp)^\flat=_\BCI \BB$,
$(\CC^\sharp)^\flat=_\BCI \CC$ and
$(\II^\sharp)^\flat=_\BCI \II$. 
}%

\begin{lemma}\label{lem:lambda-star}
$(\lambda^* x.P)^\sharp =_{\beta} \lambda x.P^\sharp$.
\end{lemma}
\comment{%
\underline{Proof}
Induction on $P$.
\begin{itemize}
\item
$P\equiv x$: 
$$(\lambda^*x.x)^\sharp\equiv\II^\sharp\equiv\lambda x.x\equiv\lambda x.x^\sharp.$$
\item
$P\equiv Q\,R$ with $x\in fv(Q)$:
$$
\begin{array}{rlll}
(\lambda^*x.Q\,R)^\sharp
&\equiv&
(\CC\,(\lambda^*x.Q)\,R)^\sharp\\
&\equiv&
\CC^\sharp\,(\lambda^*x.Q)^\sharp\,R^\sharp\\
&=_{\beta}&
\CC^\sharp\,(\lambda x.Q^\sharp)\,R^\sharp
&\mbox{i.h.}\\
&\equiv&
(\lambda xyz.x\,z\,y)\,(\lambda x.Q^\sharp)\,R^\sharp\\
&=_{\beta}&
\lambda z.(\lambda x.Q^\sharp)\,z\,R^\sharp\\
&=_{\beta}&
\lambda x.Q^\sharp\,R^\sharp\\
&\equiv&
\lambda x.(Q\,R)^\sharp
\end{array}
$$
\item
$P\equiv Q\,R$ with $x\in fv(R)$:
$$
\begin{array}{rlll}
(\lambda^*x.Q\,R)^\sharp
&\equiv&
(\BB\,Q\,(\lambda^* x.R))^\sharp\\
&\equiv&
\BB^\sharp\,Q^\sharp\,(\lambda^* x.R)^\sharp\\
&=_\beta&
\BB^\sharp\,Q^\sharp\,(\lambda x.R^\sharp)
&\mbox{i.h.}\\
&\equiv&
(\lambda xyz.x\,(y\,z))\,Q^\sharp\,(\lambda x.R^\sharp)\\
&=_\beta&
\lambda z.Q^\sharp\,((\lambda x.R^\sharp)\,z)\\
&=_\beta&
\lambda x.Q^\sharp\,R^\sharp\\
&\equiv&
\lambda x.(Q\,R)^\sharp
\end{array}
$$
\end{itemize}
}%
\begin{lemma}\label{lem:LLambda2BCI2LLambda-is-identity}
$(M^\flat)^\sharp =_{\beta\eta} M$. 
\end{lemma}
\underline{Proof}
Induction on $M$. Only the case of lambda abstraction is nontrivial,
in which we use Lemma \ref{lem:lambda-star}.

\begin{proposition}
$P=_\BCI Q$ iff $P^\sharp=_{\beta\eta}Q^\sharp$.
\end{proposition}
\underline{Proof}
$P^\sharp=_{\beta\eta}Q^\sharp$ implies
$P=_\BCI(P^\sharp)^\flat=_\BCI (Q^\sharp)^\flat=_\BCI Q$
by Lemma \ref{lem:BCI2LLambda2BCI-is-identity} and \ref{lem:eqLLambda-implies-eqBCI}.

\begin{proposition}
$M=_{\beta\eta} N$ iff $M^\flat=_\BCI N^\flat$.
\end{proposition}
\underline{Proof}
$M^\flat=_\BCI N^\flat$ implies
$M=_{\beta\eta} (M^\flat)^\sharp =_{\beta\eta} (N^\flat)^\sharp =_{\beta\eta}  N$ by Lemma \ref{lem:LLambda2BCI2LLambda-is-identity} and \ref{lem:eqBCI-implies-eqLLambda}.

\end{document}